\documentclass[preprint]{ptephy_v1}
\usepackage{graphicx}% Include figure files
\usepackage{dcolumn}% Align table columns on decimal point
\usepackage{bm}% bold math
\begin{document}

\title{Observing how deep neural networks understand physics through the energy spectrum of one-dimensional quantum mechanics}
\author{Kenzo Ogure}
\affil{Department of Nuclear Engineering, Kyoto University, Kyoto daigaku-katsura,
Nishikyo-ku, Kyoto, 615-8540, Japan\email{ogure.kenzo.8v@kyoto-u.ac.jp}}
\begin{abstract}
We investigated how neural networks(NNs) understand physics using one-dimensional quantum mechanics. After training an NN to accurately predict energy eigenvalues from potentials, we used it to confirm the NN’s understanding of physics from four different aspects. The trained NN could predict energy eigenvalues of different kinds of potentials than the ones learned, predict the probability distribution of the existence of particles not used during training, reproduce untrained physical phenomena, and predict the energy eigenvalues of potentials with unknown matter effect. These results show that NNs can learn the physical laws from experimental data, predict the results of experiments under conditions different from those used for training, and predict physical quantities of types not provided during training. Because NNs understand physics in a different way than humans, they will be a powerful tool for advancing physics by complementing the human way of understanding. 
\end{abstract}
\subjectindex{A44 Neural networks, A2 Computational physics}

\maketitle

\section{\label{sec:intro}Introduction}
In recent times, deep neural networks (DNNs) have made remarkable progress in image recognition, natural language processing, voice recognition, and anomaly detection through numerous technological breakthroughs. Among these breakthroughs, the residual connection prevents gradient loss, even when neural networks (NNs) have deep layers, contributing to the very high image recognition capability\cite{he2016deep}. Another example is the attention mechanism, which has succeeded in connecting neurons in distant locations, a shortcoming of convolutional neural networks (CNNs), and has significantly advanced fields such as translation, where relationships between distant words are essential\cite{vaswani2017attention,devlin2018bert}. In addition to accuracy improvement, NNs have begun generating new images or sentences by themselves, and their performance has been rapidly improving \cite{kingma2013auto,goodfellow2014generative}.  In a slightly different field, the combination of DNNs and reinforcement learning has rendered humans incapable of competing with computers in table games such as Go and Shogi, where human intuition was previously superior\cite{silverMastering2016,silverMastering2017}.

The significant difference between NNs and previous artificial intelligence is that NNs do not require human intuition at the design stage. Earlier types of artificial intelligence relied on humans to predetermine features of objects and perform learning. In contrast, NNs do not require such human intuition and find features in the learning process by themselves. In return, NNs comprise numerous parameters and require high computing power for learning. In addition, it is difficult for humans to understand how trained NNs operate since they do not have the predetermined features humans want to observe to understand the operation.

Another advantage of NNs is that the input and output can be exchanged in some cases, enabling their application to inverse problems.  If the inputs and outputs of the problem are bijections, they can be swapped and trained to create a NN that solves the inverse problem.

The properties of not requiring prior intuition and applicability to inverse problems make NNs promising in natural sciences, where the goal is to reveal unknown problems, and NNs have been used in many fields, including physics\cite{carleoMachine2019}. In this paper, the subject of this study is the Schr\"{o}dinger equation in quantum mechanics, and even if we focus only on the surrounding area, we can see several recent developments related to NNs. Partial differential equations, including the Schr\"{o}dinger equation, have been solved using NNs\cite{raissiPhysicsinformed2019}, potentials have been estimated inversely from wave functions\cite{hongPredicting2021,sehanobishLearning2021}, and soliton solutions to the nonlinear Schr\"{o}dinger equation have been investigated using NNs\cite{puSoliton2020}. A new family of NNs inspired by the Schr\"{o}dinger equation has also been proposed\cite{nakajimaNeural2021}.

In these studies, the Schr\"{o}dinger equation was used, and in this sense, human understanding of physics was already assumed. However, since we consider this study preparation for applying NNs to unsolved physical systems, we do not directly solve the Schr\"{o}dinger equation using NNs. From this viewpoint, the research conducted in a context most similar to this paper is Ref.\cite{millsDeep2017}. This pioneering study provided NNs with two-dimensional (2D) potentials and the corresponding energy eigenvalues of the ground states. This prior work and our study differ in technical aspects, such as the method for generating the potentials and NN’s architecture, but the most significant difference is that we are most interested in how NNs understand physics. In this sense, despite the differences in physical systems and methods, our study shares a common interest with Ref.\cite{itenDiscovering2020}.

The application of NNs to concrete physics problems seems to be progressing well. However, how an NN learns physics is embedded in numerous parameters and is difficult to find. In particular, whether the NN is merely learning its output pattern or understanding the underlying physics is essential in considering the future use of NNs. If the NN understands physical laws, it should be able to predict physical quantities of various ranges and types that were not used in training. It should also be possible to reproduce physical phenomena and estimate physical quantities even if the laws governing that physical system were not yet known. This study sheds light on these points using one-dimensional (1D) quantum mechanics as a subject. Since our research assumes that we will use NNs to solve unknown physics problems in the future, we only provide NN potentials and a few energy eigenvalues, which are physical quantities that can be measured experimentally.

This paper is organized as follows. First, in Sec.\ref{sec:method}, after a brief introduction to NNs for physicists, we demonstrate how to generate datasets and training results in this paper. Then, in Sec.\ref{sec:understanding}, where we state our main results, we investigate four aspects of how NNs understand physics and the potential for NNs to be used in physics research. Next, section \ref{sec:conclusion} summarizes the results of this paper and provides some insights into the future potential of NNs in physics. Finally, the NN architecture used in this paper is summarized in the appendix.

\section{\label{sec:method}Setup and training of neural networks}
This section first presents a brief introduction to NNs for physicists and then explains how NNs are trained in this paper on 1D quantum mechanics problems in Sec.\ref{subsec:structure}. Next, we explain how to prepare the datasets in Sec.\ref{subsec:dataset}, and finally, we present the training results in Sec.\ref{subsec:learning}.

\subsection{\label{subsec:structure}Application of Neural Networks to one-dimensional Quantum Mechanics}
Regression using an NN is a kind of variational method if we use a term familiar to physicists. An NN is a function that contains parameters, which are determined to reproduce the best ground truth outputs. This function, inspired by the brain’s structure, has several characteristics:

\begin{itemize}
    \item 
    While the variational trial function in physics is typically assumed to be an elementary function or a superposition of elementary functions by some physical intuition, NNs do not require such intuition. In return, NNs have far more parameters than trial functions in traditional variational methods in physics to be sufficiently flexible.
    \item
    Since NNs are developed by a composite function of linear transformations and simple functions called activation functions, the back-propagation technique can quickly obtain their derivatives.
    \item
    As a technical matter, NNs are beginning to be firmly recognized for their usefulness, and supportive libraries such as Pytorch and Tensorflow have been developed, as well as technologies such as Cuda for faster computation on GPUs. This has enabled experimenting with NN technology to solve different problems easily.
\end{itemize}

Here, we explain the training of NNs in more detail. Training an NN requires a dataset recorded through experience, experimentation, or artificial data generation. We write this as $\{(v_m^{(k)},E_i^{(k)})\}_{k=1,\cdots, N}$, where $v_m^{(k)}\ (m=0,\cdots,M)$ is real-valued input vectors of dimension $M+1$, and $E_i^{(k)}\ (i=0,\cdots,i_{max}-1)$  is real-valued output vectors of dimension $i_{max}$. The index $k$ is the sample's label, which can take values from $1$ to $N$. The dataset size $N$ should be as large as possible to train NNs.

An NN can be written as $f_i(v_m^{(k)};\{w_{\alpha}\})$. In addition to taking $v_m^{(k)}$ as an argument, it has real valued parameter set $\{w_{\alpha}\}$ with index $\alpha$, which can take values from 1 to the total number of the NN's parameters. The majority of $\{w_{\alpha}\}$ is the weights and bias of the linear combination used to create the composite function, but other parameters can also be included. Training an NN means optimizing $\{w_{\alpha}\}$ so that $f_i(v_m^{(k)};\{w_{\alpha}\})$ is as close to $E_i^{(k)}$ as possible for most of $v_m^{(k)}$.

\begin{figure}
\includegraphics[width=165mm]{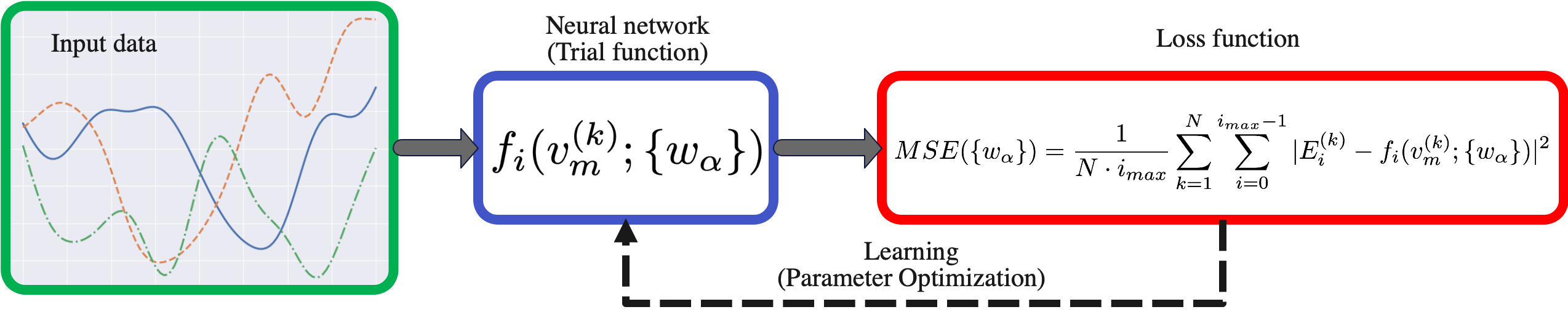}
\caption{\label{fig:flow}Training a neural network(NN). An NN is a function, which takes a multidimensional real-valued input and produces a multidimensional real-valued output. Training an NN requires many input and ground truth output datasets. NN learning is the optimization of real valued parameter set, $\{w_{\alpha}\}$, to reduce the loss function, $MSE(\{w_{\alpha}\})$,  linked to the difference between the correct data and the prediction by the output of the NN. In this optimization, the back-propagation method, which can quickly calculate the derivative of the loss function to parameters, is indispensable. In this paper, the input is 1D potentials’ values on mesh points, and the output is the corresponding energy spectra. We describe the architecture of our NN and the specific method for optimizing its parameters in the appendix.}
\end{figure}

The flow of this optimization is depicted in Fig.\ref{fig:flow}. First, when the input data $v_m^{(k)}$ is fed to the NN, the output $f_i(v_m^{(k)};\{w_{\alpha}\})$ is obtained. Then, the loss resulting from the error between the output and the ground truth data, $E_i^{(k)}$, is calculated. In addition, the first-order derivative of the loss function to each parameter, $w_{\alpha}$, can be calculated almost simultaneously using a method called back-propagation. The value of $w_{\alpha}$ is updated to decrease the loss using the derivative. Repeating this process until the loss does not significantly decrease is called NN training.

In actual training, instead of adding up all samples in the loss function, a few samples are selected, which is called mini-batch learning. Mini-batch learning saves computer memory, and it is empirically known that mini-batch learning is faster than taking all data. Furthermore, the final results are often better when the batch size is smaller. At least, we do not know the general theory for determining the optimal batch size and have empirically fixed the size to 100 by trial and error in this paper.

Next, we explain how we apply NN learning to a physical problem in this paper. We deal with quantum mechanics in one dimension. A particle is confined to the interval $-1<x<1$ by infinitely high walls. Then, we generate $N$ various potentials between $-1<x<1$ and take their values at equally spaced points as input data $v_m^{(k)}$. For the output data $E_i^{(k)}$, we take ten of the smallest energy eigenvalues corresponding to the potentials ($i_{max}=10$). The following subsection explains how to generate the potentials and calculate the corresponding eigenvalues. In this study, $E_i^{(k)}$ is obtained by solving the Schrödinger equation numerically. However, because the primary goal of this paper is to see if NN can be trained to understand the physical laws using observed physical quantities, $E_i^{(k)}$ is positioned as simulated experimental data.

We use a 1D CNN. In addition, we also use the residual connection and the self-attention layer, which are effective when used in conjunction with CNNs in the natural language processing and image generation fields\cite{he2016deep,zhang2019self}. We adopt the mean squared error(MSE) between the ground truth energy eigenvalues and the outputs from the NN for the loss function. We describe the architecture of our NN and the specific method for optimizing its parameters in the appendix.

\subsection{\label{subsec:dataset}How to prepare the datasets}
We take the energy spectrum of a particle confined to an interval in one dimension as our output data. Hence, we consider the following potential,
\begin{eqnarray}
V^{(k)} (x)= \left\{
\begin{array}{ll}
v^{(k)}(x) & (|x| \leq 1)\\
\infty & (|x| > 1)
\end{array}
\right.
\label{eq:potential}
\end{eqnarray}
where $v^{(k)}(x)$ are arbitrary continuous functions and k takes values from 1 to N, which is the size of datasets.

Next, we use the kernel method to generate the potential functions, $v^{(k)}(x)$. Note that we use the kernel method only to generate the potential functions, not to perform regression using the Gaussian process. 
We set mesh points on $-1 <  x < 1$ with $\Delta x\  (=2/M)$ width as $x_l = (2l-M)/M \ (l = 0,1,2,\cdots M)$ to use the kernel method. Let the kernel function be $k(x,x^{'})$, and then, under our discretization, the kernel matrix is the $(M+1)\times (M+1)$ matrix defined by $K_{lm}=k(x_l,x_m)$. By using this kernel matrix as the covariance matrix, we can probabilistically generate the potential function, as follows:
\begin{equation}
{\bm v} \sim {\mathcal{N}}({\bf 0},{\bf K})
\end{equation}
where $\mathcal{N}$ represents the normal distribution function, with the mean being ${\bf 0}$ and the covariance matrix being ${\bf K}$. Furthermore, ${\bm v}$ is an $(M+1)$-dimensional vector, whose components are regarded as potential values at $x_l$.  We obtain ${v}^{(k)}(x_l)$ by performing this potential generation for $N$ times.

\begin{figure}
\includegraphics[width=165mm]{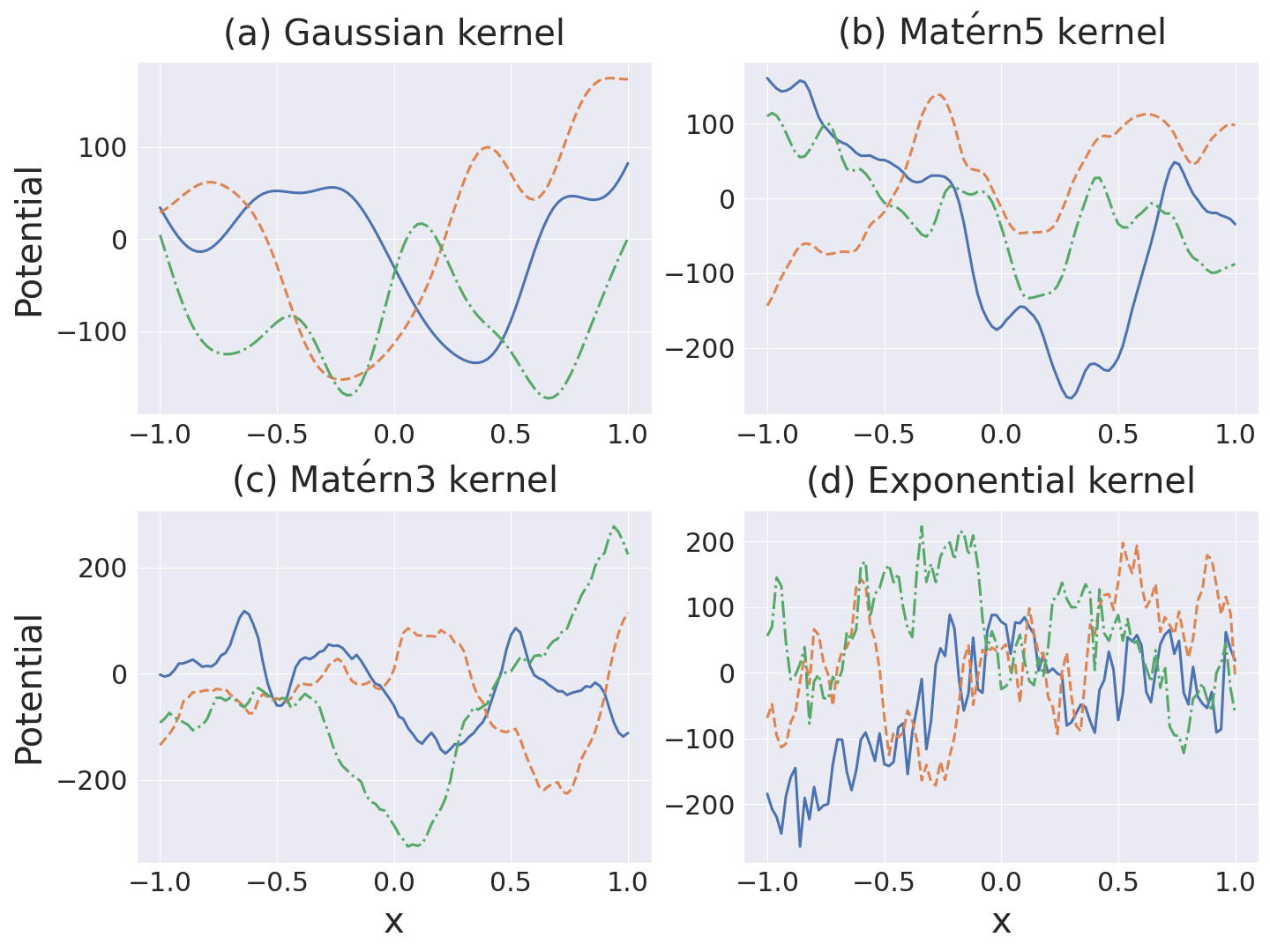}
\caption{\label{fig:potential}Potentials generated using the kernel method. The kernel parameters were taken to be the correlation distance $L = 0.2$ and the amplitude $\sigma= 100$ for all kernels. The smoothness varies depending on the kernel used. (a) Potentials generated from the Gaussian kernel, Eq.(\ref{eq:gauss}) or Eq.(\ref{eq:knu}) at $\nu = \infty$. These are infinite-order differentiable. (b) Potentials generated from the Mat\'ern5 kernel, Eq.(\ref{eq:m5}) or Eq.(\ref{eq:knu}) at $\nu = \frac{5}{2}$. These are second-order differentiable. (c) Potentials generated from the Mat\'ern3 kernel, Eq.(\ref{eq:m3}) or Eq.(\ref{eq:knu}) at $\nu = \frac{3}{2}$. These are first-order differentiable. (d) Potentials generated from the exponential kernels, Eq.(\ref{eq:exp}) or Eq.(\ref{eq:knu}) at $\nu = \frac{1}{2}$. These are not differentiable.}
\end{figure}

This study uses four kernel functions to generate the training data: Gaussian kernel $k^{(RBF)}(x,x^{'})$, Mat\'ern5 kernel $k^{(M5)}(x,x^{'})$, Mat\'ern3 kernel $k^{(M3)}(x,x^{'})$, and exponential kernel $k^{(Exp)}(x,x^{'})$.  These functions are defined, as follows:
\begin{eqnarray}
k^{(RBF)}(x,x^{'}) &=& \sigma^2 \exp{\left(-\frac{r^2}{2L^2}\right)}\label{eq:gauss}\\
k^{(M5)}(x,x^{'}) &=& \sigma^2 \left(1+\frac{\sqrt{5}r}{L}+\frac{5r^2}{3L^2}\right)\exp{\left(-\frac{\sqrt{5}r}{L}\right)}\label{eq:m5}\\
k^{(M3)}(x,x^{'}) &=& \sigma^2 \left(1+\frac{\sqrt{3}r}{L}\right)\exp{\left(-\frac{\sqrt{3}r}{L}\right)}\label{eq:m3}\\
k^{(Exp)}(x,x^{'}) &=& \sigma^2 \exp{\left(-\frac{r}{L}\right)}\label{eq:exp}
\end{eqnarray}
where we defined $r=|x-x^{'}|$. These functions have two parameters: $\sigma$, which controls the magnitude of the generated potential, and $L$, which controls the correlation distance of the potential. These kernel functions can be written concisely using the gamma function $\Gamma (\nu)$, and the Bessel function of the second kind $K_{\nu}$, as follows:
\begin{eqnarray}
k_{\nu}(x,x^{'}) &=& \frac{2^{1-\nu}\sigma^2}{\Gamma (\nu)}\left(\frac{\sqrt{2\nu}r}{L}\right)^{\nu}K_{\nu}\left(\frac{\sqrt{2\nu}r}{L}\right)\label{eq:knu}
\end{eqnarray}
The Gaussian, Mat\'ern5, Mat\'ern3, and the exponential kernels correspond to the cases of $\nu = \infty,\ \frac{5}{2},\ \frac{3}{2},\ \frac{1}{2}$, respectively. Kernels with larger $\nu$ generate smoother potential functions, and the generated potentials are $[\nu]$ times differentiable. Here  $[\nu]$ is the largest integer not greater than $\nu$.  Figure \ref{fig:potential} shows the potentials generated from each kernel.  The four panels in this figure show that the smoothness of the potentials changes as $\nu$ increases.

Now that we have generated the potentials, which are the input data needed to train the NN, we need to find the energy spectrum of those potentials, the output data. For this purpose, we solve the time-independent 1D Schr\"{o}dinger equation,
\begin{equation}
\left\{-\frac{1}{2} \frac{d^2}{dx^2}+V^{(k)}(x)\right\} \psi^{(k)}(x)=E^{(k)} \psi^{(k)}(x),\label{sch-eq}
\end{equation}
where $\psi^{(k)}(x)$ is the wave function and we use Planck's constant $h = 2\pi$, and the particle's mass $m=1$. Since we are considering the potentials in Eq.(\ref{eq:potential}), the boundary conditions for the wave function are $\psi^{(k)}(\pm 1)=0$. In the case of the square potential, $v^{(k)}(x) = 0$, the energy eigenvalues are,
\begin{equation}
E_{i}^{(k)}=\frac{\pi^2 (i+1)^2}{8} \ \ \  (i=0,1,2,\cdots),\label{eq:square_energy}
\end{equation}
and the corresponding wave functions are,
\begin{equation}
\psi^{(k)}_{i}(x)=\sin{\left[\frac{\pi}{2}(i+1)(x+1)\right]}.
\label{eq:cwave}
\end{equation}
Since we use the ten lowest energy eigenvalues as the ground truth output data, it should have a rich structure.  Referring to Eq.(\ref{eq:square_energy}), we take $\sigma = 100$ in Eqs.(\ref{eq:gauss})-(\ref{eq:exp}) throughout this paper expect the dataset for obtaining Fig.\ref{fig:generalization_s6}. Figure \ref{fig:learning_result} shows that this setting works in the actual numerical experiment results.

We use the matrix method\cite{izaacComputational2019a} to solve this eigenvalue problem for general $v^{(k)}(x)$. This method is appropriate for our setting because it enables us to find multiple energy eigenvalues and corresponding wave functions simultaneously. Furthermore, since we have already discretized the problem to generate the values of the potentials on the mesh points, we can directly use this method.

Applying the matrix method to Eq.(\ref{sch-eq}) is straightforward: rewriting Eq.(\ref{sch-eq}) by replacing the second-order derivative with a difference results in the following matrix eigenvalue problem,
\begin{eqnarray}
H^{(k)}\Psi^{(k)} =E^{(k)}\Psi^{(k)}\label{eq:equation_of_motion}
\end{eqnarray}
which we defined, as follows ($v_m^{(k)} \equiv v^{(k)}(x_m)$),
\begin{eqnarray}
H^{(k)}
=
-\frac{1}{2(\Delta x)^2}\left(\begin{array}{cccccc}
-2 & 1 & 0 & \ldots & 0 & 0 \\
1 & -2 &  1 & \ldots & 0 & 0 \\
0 &  1 & -2 & \ldots & 0 & 0 \\
\vdots & \vdots & \vdots & \ddots & \ddots & \vdots \\
0 & 0 & 0 & \ddots & -2 &  1 \\
0 & 0 & 0 & \ldots &  1 & -2
\end{array}\right)
+
\left(\begin{array}{cccccc}
v^{(k)}_1 & 0 & 0 & \ldots & 0 & 0 \\
0 & v^{(k)}_2 &  0 & \ldots & 0 & 0 \\
0 &  0 & v^{(k)}_3 & \ldots & 0 & 0 \\
\vdots & \vdots & \vdots & \ddots & \ddots & \vdots \\
0 & 0 & 0 & \ddots & v^{(k)}_{M-2} &  0 \\
0 & 0 & 0 & \ldots &  0 & v^{(k)}_{M-1}
\end{array}\right)\label{eq:hamiltomian}
\end{eqnarray}
Here, we have used the boundary conditions $\psi^{(k)}(x_0) = \psi^{(k)}(x_M) = 0$, so the matrix $H^{(k)}$ is an $(M-1)\times (M-1)$ dimensional matrix and the eigenvector $\Psi^{(k)}$ is $(M-1)$-dimensional vector. This eigenvalue problem is easy to solve because $H^{(k)}$ is a tridiagonal real symmetric matrix.  We denote the eigenvectors and corresponding eigenvalues as $E_i^{(k)}$ and
\begin{eqnarray}
\Psi^{(k)}_i
&=&
\left(
\begin{array}{c}
\psi^{(k)}_i(x_1) \\
\psi^{(k)}_i(x_2)  \\
\psi^{(k)}_i(x_3)  \\
\vdots \\
\psi^{(k)}_i(x_{M-2})  \\
\psi^{(k)}_i(x_{M-1})  
\end{array}
\right),\hspace{1cm}
\Psi^{(k)T}_i\Psi^{(k)}_i = 1\label{eq:wave-nomalization}
\end{eqnarray}
, respectively. Note that this normalization of the wave function is natural as a normalization for the matrix eigenvalue problem but deviates by $\sqrt{\Delta x}$ from the ordinary normalization condition of the wave function, $\int_{-1}^{1}|\Psi^{(k)}(x)|^2 dx=1$. There are M-1 eigenvalues, labeling them as $E_i^{(k)} < E_j^{(k)}$ for $i<j$. Since this is 1D quantum mechanics, the energy eigenvalues have no degeneracy. We adopt ten eigenvalues($i=0,1,2\cdots,9$) for the ground truth output data of the NN.

We have depicted one potential generated from the Gaussian kernel and the ten lowest energy eigenvalues of the potential in Fig.\ref{fig:learning_result}(c). We note that we have confirmed that these are the ten lowest energy eigenvalues by counting the nodes of the corresponding wave functions.

This subsection explains how to create a set of ($M+1$)-dimensional potential vectors, $v_m^{(k)}$, as input data and 10-dimensional energy eigenvalues, $E_i^{(k)}$, as output data for training an NN. Here we used the Schr\"{o}dinger equation to prepare these datasets, but we did not feed the equation to the NN. We only feed the NN with the set of potentials $v_m^{(k)}$ and energy eigenvalues  $E_i^{(k)}$ as if we observe them in a physical experiment. Of course, the Schr\"{o}dinger equation is not required if the datasets are prepared by conducting experiments to measure energy eigenvalues while varying the potential in various ways. The primary goal of this paper is to demonstrate that NN understands physical laws using only the dataset $\{(v_m^{(k)},E_i^{(k)})\}_{k=1,\cdots, N}$ principally observed in the experiment.

\subsection{\label{subsec:learning}Learning setup and results}
We generated three training datasets, N1, N2, and N5 with the sizes, $1\times 10^5$, $2\times 10^5$, and $5\times 10^5$, respectively, for all four kernels of Eq.(\ref{eq:gauss})-(\ref{eq:exp}) to train the NN. The reason why we constructed datasets of different sizes was to examine the degree of overfitting. In addition, we generated different datasets of sizes $1\times 10^4$ for all kernels for validation. The parameters of the kernels were $\sigma = 100$ and $L=0.2$. Since the batch size was fixed at 100, one epoch of training would update the NN parameters $\{w_{\alpha}\}$, $1\times 10^3$, $2\times 10^3$, and $5\times 10^3$ times for the datasets  N1, N2, and N5, respectively. To keep the total number of parameter updates constant and compare the learning status fairly, we trained $100$, $50$, and $20$ epochs for the datasets N1, N2, and N5, respectively. This setup resulted in the parameter set, $\{w_{\alpha}\}$, being updated $1\times 10^5$ times in total for any dataset. The loss function is the MSE. The design of the NN and the precise optimization method of the parameter set,  $\{w_{\alpha}\}$, are included in the appendix.

Figure \ref{fig:learning_result}(a) shows the learning progress. Here, we defined the normalized mean squared error(NMSE) to evaluate the error as a dimensionless quantity, as follows,
\begin{equation}
    NMSE(\{w_{\alpha}\})=
    \frac{\sum_{k=1}^{\cal N}\sum_{i=0}^{i_{max}-1}(E_i^{(k)}-f_i(v_m^{(k)};\{w_{\alpha}\}))^2}{\sum_{k=1}^{\cal N}\sum_{i=0}^{i_{max}-1}(E_i^{(k)})^2}.\label{normalized_mse}
\end{equation}
Here, ${\cal N}$ is the size of the used dataset. In addition, we defined the progress rate as the ratio of the performed update times of $\{w_{\alpha}\}$ to the total one, $1\times 10^5$. This figure displays the NN’s learning process using datasets generated from the Gaussian kernel. The dashed line represents the error for the training data, and the solid line represents the error for the validation data. This figure shows that overfitting is suppressed as the size of the dataset increases, and the effect of overfitting is minimal at N5. The NMSE is less than $10^{-4}$ at the end of the training, and the error in the energy eigenvalue is less than $1\%$ on average for N5. We have confirmed the datasets generated from other kernels show the same trend even though some quantitative differences exist.

Figure \ref{fig:learning_result}(b) shows the difference in the NMSE after the completion of training depending on the kernel that generates the training data. In the figure legend hereafter, Gaussian, Mattern 5, Mattern 3, and the exponential kernel are abbreviated as G, M5, M3, and Exp, respectively. Validation data were generated from the same kernel that generated the training data. This figure shows that the error is minor for the dataset with a larger size. In addition, the smoother the potential, the smaller the error.

Figure \ref{fig:learning_result}(c) shows  the prediction of the energy eigenvalues of one potential generated from the Gaussian kernel by an NN trained on the N5 dataset generated from the same kernel. This figure shows that the deviation of the prediction of the energy eigenvalue is small.

\begin{figure}
\includegraphics[width=165mm]{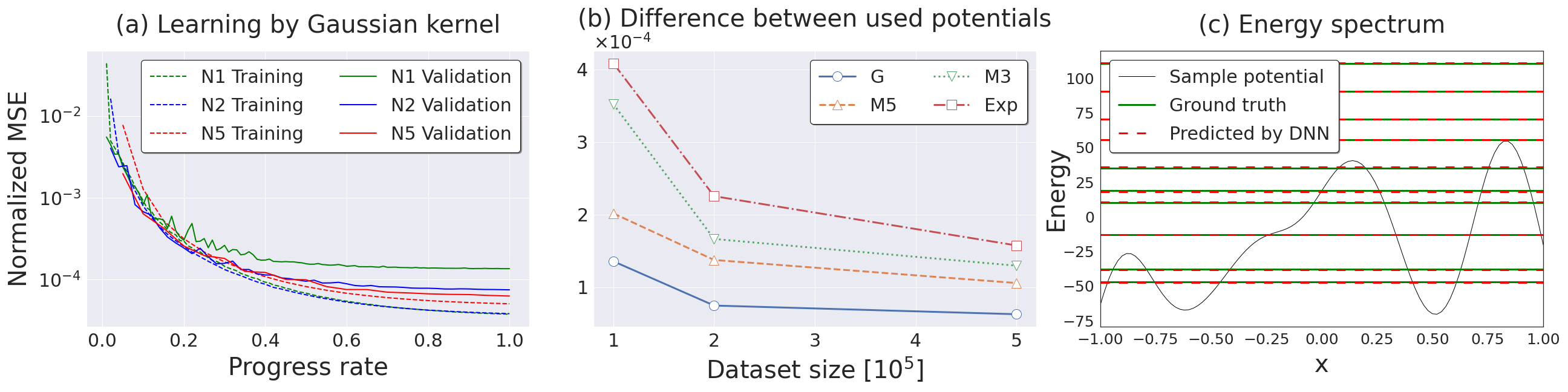}
\caption{\label{fig:learning_result}Training NNs. (a) NMSE in Eq.(\ref{normalized_mse}) is plotted against the training progress. The dataset used for training varies between $1\times10^5$(N1), $2\times10^5$(N2), and $5\times10^5$(N5) to study the difference in learning progress. The total number of parameter updates is $1\times10^5$ in all cases. The dashed line represents the error for the training data, and the solid line represents the error for the validation data. For validation, we used data generated from the same kernel that generated the data used for training. This figure shows that the larger the dataset, the less overfitting there is. (b) The final epoch errors are displayed for each kernel that generates data. This figure shows that smooth data are easier to learn. (c) For one sample potential generated from the Gaussian kernel, we compared the energy eigenvalues predicted by the NN with the ground truth values calculated from the matrix diagonalization in Eq.(\ref{eq:equation_of_motion}). As shown in (b), for the data generated from the Gaussian kernel, the average error of the energy eigenvalues is less than $1\%$, which can barely be seen in this figure.}
\end{figure}

\section{\label{sec:understanding}Understanding Physics with Neural Networks}
In Sec.\ref{sec:method}, we showed that NNs can now predict energy eigenvalues accurately, but it is still unclear whether they understand the physics described by Eq.(\ref{sch-eq}). As described in the Appendix, our NN has $4.9\times 10^6$ real number parameters, $\{w_{\alpha}\}$, and the real number size of the output of the largest dataset, $\{E_i^{(k)}\}_{k=1,\cdots,N}$, is $5\times 10^6$. This fact may lead us to suspect that the NN does not understand physics but memorizes the output.

In this section, which describes the main results of this paper, we will show in four aspects that NNs do understand physics and that they are helpful for the study of physics. First, in Sec.\ref{subsec:generalization}, we show that NNs can predict energy eigenvalues even for potentials of a different shape than the training dataset. Next, in Sec.\ref{subsec:potential}, we successfully extract the eigenvectors' magnitudes not used for the training from the trained NN. We then show that NNs can predict unknown physical phenomena in Sec.\ref{subsec:crossl}.  Finally, we treat potentials with matter effects unknown to the experimenters to demonstrate the broad applicability of the method in Sec.\ref{subsec:adaptability}.

These results show that NNs can learn the physical laws from experimental data, predict the results of experiments under conditions different from those used for training, and predict physical quantities of types not provided during training. Because NNs understand physics in a different way than humans, they will be a powerful tool for advancing physics by complementing the human way of understanding. 

\subsection{\label{subsec:generalization}Expand the scope of applicable data}
In Sec.\ref{subsec:learning}, the training performance of the NN was verified with validation data generated from the same kernel that generated the training data. However, if the NN understands the physics governed by Eq.(\ref{sch-eq}), it should predict the energy eigenvalues of a potential generated from a different kernel than the one that generated the training data. Therefore, this subsection shows how the NN predicts energy eigenvalues for the validation data generated from a different kernel than the training data generated.

\begin{figure}
\includegraphics[width=165mm]{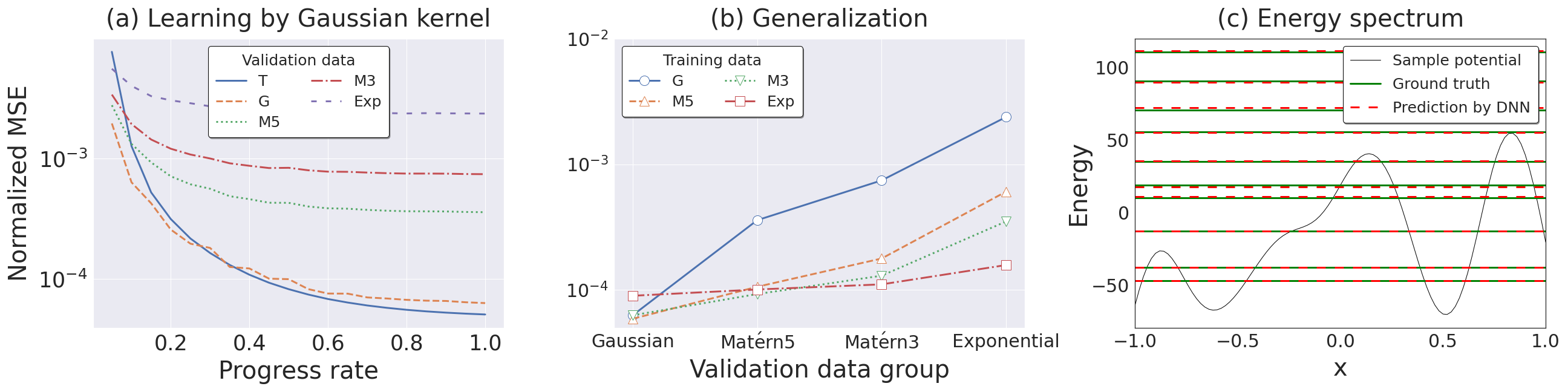}
\caption{\label{fig:generalization}(a) NMSE versus learning progress when using data generated from the Gaussian kernel for training dataset. The predictions for validation datasets generated from the other kernels are also good, and it seems that predictions for smooth potentials are easier than rough ones. (b) The error at the final epoch is shown for each N5 dataset used for training. The NN trained with the exponential kernel shows a good prediction for any validation dataset. (c) The NN trained by the N5 dataset generated by the exponential kernel was used to predict energy eigenvalues for the same potential as in Fig.\ref{fig:learning_result}(c). The energy eigenvalues of the seventh excited state are slightly different, but considering that the prediction is for the potential generated by the the different kernel from the training one, the prediction is excellent.}
\end{figure}

Figure \ref{fig:generalization}(a) shows the training progress using the N5 dataset generated from the Gaussian kernel. This figure shows that the error for the validation data generated from the Gaussian kernel is the smallest, but the prediction is not broken, even for datasets generated from other kernels.

We show the error at the final epoch in Fig.\ref{fig:generalization}(b)  by the kernel that generated the training data. This figure shows that NNs trained with data such as the one in Fig.\ref{fig:potential}(d), generated from the exponential kernel, show promising results, regardless of the dataset type. On the other hand, the NN trained on the dataset generated from the Gaussian kernel is not good at predicting potentials that are not smooth. 

We used the NN trained on the N5 dataset generated from the exponential kernel in Fig.\ref{fig:generalization}(c) to predict the energy eigenvalues of the same potential as in Fig.\ref{fig:learning_result}(c). This figure demonstrates that, although there is a slightly significant deviation in the energy eigenvalue of the seventh excited state, the overall prediction is accurate.  Though the NN has seen only the jagged potential in Fig. \ref{fig:potential}(d), it correctly predicted the energy eigenvalues for a smooth potential generated from Gaussian kernel. This result indicates that it understood the physical laws governing this system.

\begin{figure}
\includegraphics[width=165mm]{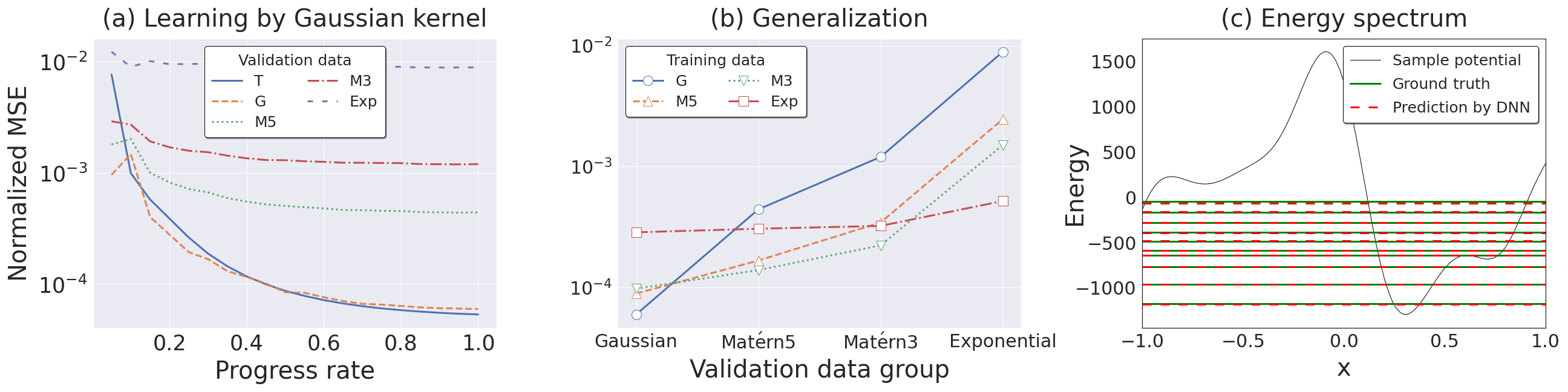}
\caption{\label{fig:generalization_s6}The same figures as in Fig. \ref{fig:generalization} for the dataset, N5'($\sigma = 1000$).}
\end{figure}

In addition, to test whether the learning works even with highly undulating potentials, we took $\sigma = 1000$ and generated a new dataset, N5', with $5\times 10^5$ potentials for each kernel. The same figures as in Fig. \ref{fig:generalization} are shown in Fig. \ref{fig:generalization_s6}  for this new data set. As can be seen from Fig. \ref{fig:generalization_s6}(c), the ten lower energy eigenvalues taken are well below the highest point of the potentials. Figure \ref{fig:generalization_s6}(a) and (b) show that although the learning is well advanced, the error is larger than in the case of N5. This error increase is most likely due to the differences in larger energy eigenvalues; in N5, some of the larger energy eigenvalues were approaching the obvious ones determined only by the boundary conditions, but in N5', all ten energy eigenvalues are nontrivial.

Following that, we quantitatively examine the relationship between the similarity and loss between training and validation potentials. The potentials generated by different kernels appear to have different shapes in Fig.\ref{fig:potential}(a) through \ref{fig:potential}(d). However, some of the $5 \times 10^5$ potentials in N5 may be close by chance even for different kernels, and the NNs may simply remember them.

To discuss this, we define a measure of similarity between a training dataset, $\{v^{(k^{'})}_m\}_{k^{'}=1,\cdots,N}$ , and a validation potential, $u^{(k)}_m$, as
\begin{eqnarray}
d(u^{(k)}_m,\{v^{(k^{'})}_m\}_{k^{'}=1,\cdots,N})
&\equiv&
\min_{k^{'}} \min_{c_{k^{'}}} \sqrt{\frac{1}{M+1}\sum_{m=0}^{M}(u^{(k)}_m-v^{(k^{'})}_m-c_{k^{'}})^2}\\
&=&
\min_{k^{'}} 
\sqrt{\frac{1}{M+1}\sum_{m=0}^{M}(u^{(k)}_m-v^{(k^{'})}_m)^2
-{\bar \Delta}_{k^{'}}^2}\label{eq:measure}
\end{eqnarray}
, where $c_{k^{'}}$ is taken at
\begin{eqnarray}
{\bar \Delta}_{k^{'}}=\frac{1}{M+1}\sum_{m=0}^{M}(u^{(k)}_m-v^{(k^{'})}_m)\label{eq:minimum}
\end{eqnarray}
for the minimization.

\begin{figure}
\includegraphics[width=165mm]{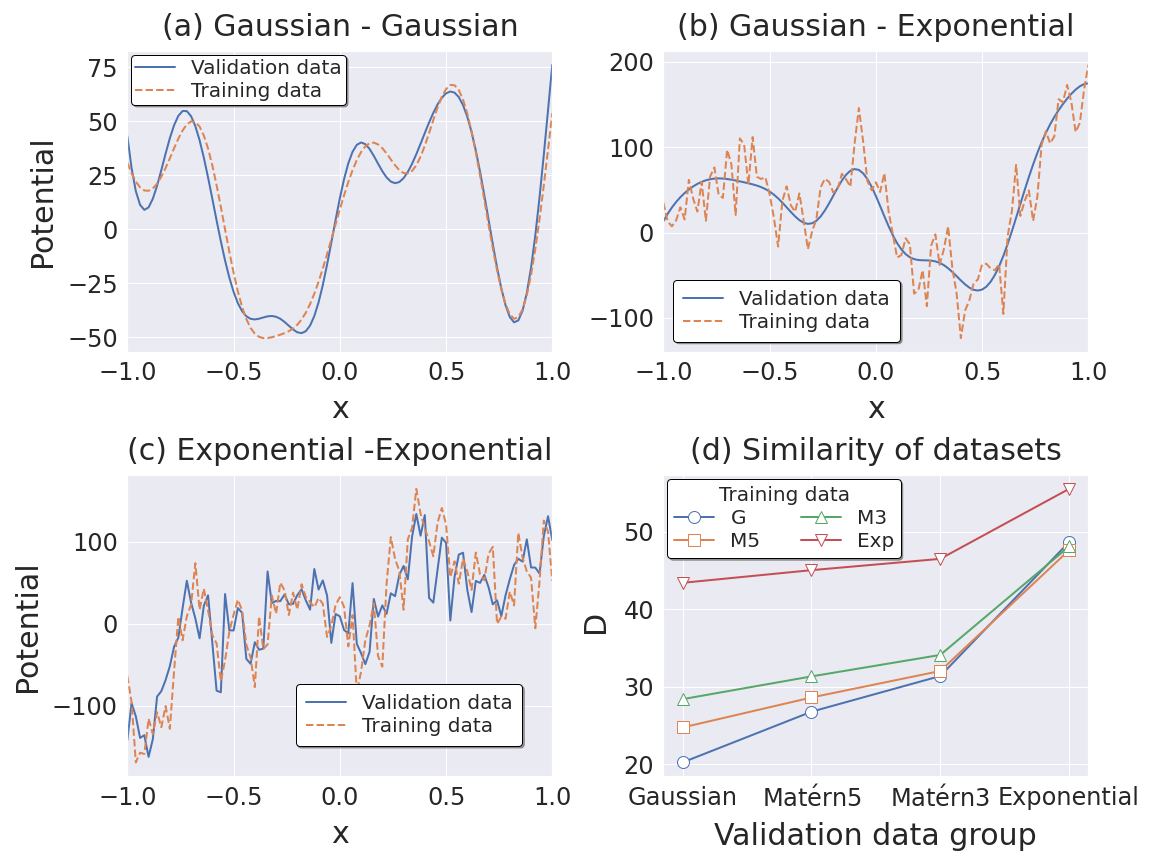}
\caption{\label{fig:potential_distance} (a) The closest potentials between validation and training dataset both generated from Gaussian kernel. (b) The potentials with the highest similarity between the validation and training datasets generated by the Gaussian and exponential kernels, respectively. (c) The exponential kernel generated the closest potentials between the validation and training datasets. (d) The resemblance of the validation and training datasets.}
\end{figure}

The smaller this number, $d$, the closer the validation potential is to the training dataset. In this definition, we find the potential with the smallest $L^2$ norm to the validation potential in the training dataset and use the norm as our measure. However, because we are concerned with the shape of the potentials, we add a constant, $c_{k^{'}}$, to the training dataset potential to minimize the $L^2$. For the validation dataset, $10^4$ potentials were generated from the Gaussian kernel, of which the one that minimizes $d$ is shown in Fig \ref{fig:potential_distance}(a). We used the N5 dataset generated from the Gaussian kernel for the training dataset. The potential that gives the minimum $d$ value in the training dataset, $v^{(k^{'})}_m+{\bar \Delta}_{k^{'}}$ is also shown. In other words, this figure shows our measure's closest pair of the $10^4$ validation potentials and the $5\times 10^5$ training potentials both generated from Gaussian kernel. We can see from this figure that if some of the validation and training data are similar in shape, we can successfully find them with our measure. Similarly, Fig.\ref{fig:potential_distance}(b) depicts the best match between the $10^4$ Gaussian kernel validation potentials and the N5 training dataset generated by the exponential kernel. However, it can be seen that their overall undulations are similar, but their shapes are not. The closest match between the validation and N5 training potentials generated by the exponential kernel is shown in Fig.\ref{fig:potential_distance}(c). This figure also demonstrates that while their overall undulations are similar, their shapes are not.

To evaluate the similarity between the validation datasets and the training datasets, the following metric is defined,
\begin{eqnarray}
D(\{u^{(k)}_m\}_{k=1,\cdots,N_v},\{v^{(k^{'})}_m\}_{k^{'}=1,\cdots,N})
=
\frac{1}{N_v}
\sum_{k=1}^{N_v}
d(u^{(k)}_m,\{v^{(k^{'})}_m\}_{k^{'}=1,\cdots,N}).
\label{eq:D}
\end{eqnarray}
Here, we take $N_v=10^4$ and $N=5\times 10^5$.

\begin{figure}
\includegraphics[width=165mm]{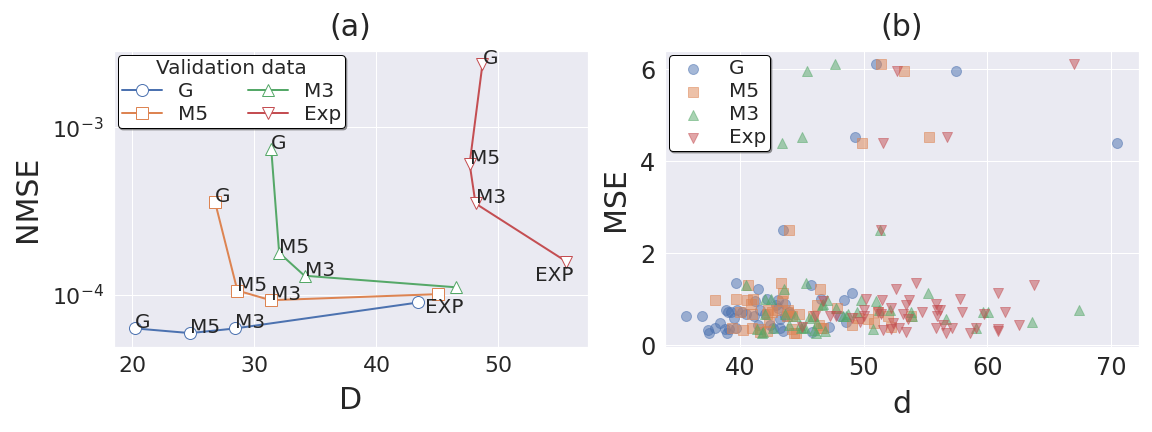}
\caption{\label{fig:loss_distance}
(a) NMSE in Fig \ref{fig:generalization}(b) are shown on the metric, $D$.  The annotations denote the datasets used for training. This figure shows that the validation dataset can have a lower loss when trained by a different kernel with a larger $D$. (b) Scatter plots of validation potentials on MSE for the ten energy eigenvalues and the measure, $d$. For training, the N5 exponential kernel dataset is used. This graph demonstrates that a lower $d$ does not always imply a lower loss.
}
\end{figure}

Figure \ref{fig:potential_distance}(d) depicts the similarity of each dataset using this metric. This figure shows that the training dataset generated by the exponential kernel is not even similar to the validation dataset generated by the exponential kernel itself. However, this diversity of potentials, prevents over-fitting and is most likely one of the reasons why the training dataset generated by the exponential kernel can train the NN to make good predictions for any of the validation data. Next, the relationship between the similarity and the loss is shown in Fig.\ref{fig:loss_distance}. Figure \ref{fig:loss_distance}(a) redraws Fig.\ref{fig:generalization}(b) with $D$ on the axis. Here, the annotations show the used training datasets. This graph demonstrates that a lower $D$ does not imply a lower loss. Instead, the validation datasets generated by Gaussian, Mat\'ern 5, and Mat\'ern 3 kernels have the lowest loss in the training datasets generated by the Mat\'ern 5, Mat\'ern 3, and exponential kernels, all of which have larger $D$ than the original kernels.

Figure \ref{fig:loss_distance}(b) shows the scatter plots of MSE for the ten eigenvalues value and $d$ for the individual validation potentials. The training dataset is N5, which was generated by the exponential kernel. This graph also demonstrates that a lower d does not always imply a lower loss. These findings suggest that while NNs do not remember potential shapes and the corresponding energy eigenvalues, they do understand the law, Eq.(\ref{sch-eq}).

Finally, we look at how the residual connection and the self-attention mechanism affect NN predictions. For this examination, we omit portions of the NN layers and observe how the loss changes. Figure \ref{fig:whole} summarizes our NN architecture. To begin, to see the effect of deepening NN via the residual connection, we examine how the loss changes as the depth of the NN changes. 

\begin{figure}
\includegraphics[width=165mm]{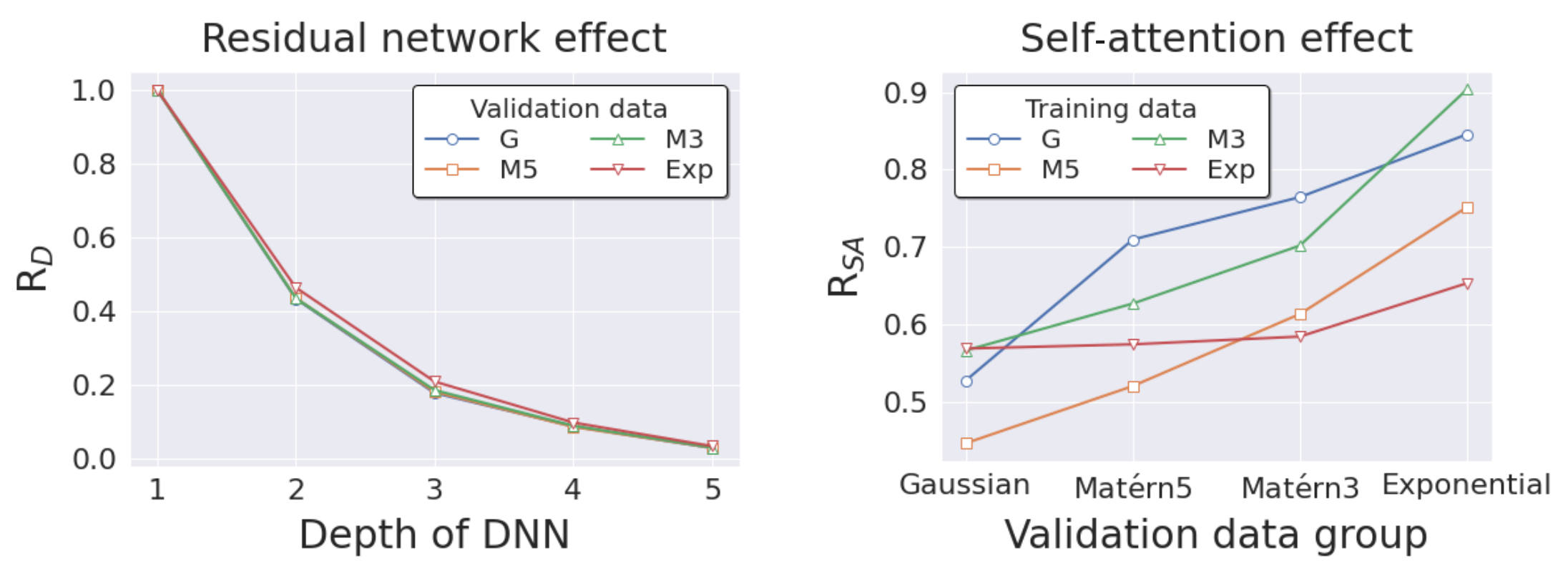}
\caption{\label{fig:resnet_sa_effect}
The left figure shows the mean loss ratio, $R_D$, defined in Eq.(\ref{eq:R_D}) on the number of used Res-SA blocks in the NN. It can be seen that by deepening the NN layers, the loss can be significantly reduced. The right figure depicts the loss ratio between NMSE with and without the self-attention mechanism, $R_{SA}$, defined in Eq.(\ref{eq:R_SA}) for all training and validation dataset combinations. The self-attention mechanism is especially useful for smooth validation data.
}
\end{figure}

As shown in Fig.\ref{fig:whole}, our NN has five blocks with residual connections, which we refer to as "Res-SA blocks." To see the difference when we reduce these blocks from the higher level, we define the following quantity, 
\begin{eqnarray}
R_D
=
\frac{\text{NMSE( Number of Res-SA blocks })}{\text{NMSE(1)}}
\label{eq:R_D}
\end{eqnarray}
, where NMSE(1) is NMSE resulting from the NN design that directly leads from the initial Res-SA block to the Flatten layer. The loss rapidly decreases as the NN is deepened using the residual connection for any training dataset, as shown in the left panel of Fig.\ref{fig:resnet_sa_effect}. This change shows that it is essential to capture the overall shape of the potential by broadening the field of view of neurons.

Next, we see the effect of the self-attention mechanism: the self-attention layer is in the Res-SA block in Fig.\ref{fig:whole}, and the contents of the Res-SA block are shown in Fig \ref{fig:res-sa}. The ratio of the NMSE evaluated from the NN with and without the self-attention layer is defined as follows,
\begin{eqnarray}
R_{SA}
\equiv
\frac{\text{NMSE( With self-attention mechanism} )}{\text{NMSE( Without self-attention mechanism )}}.
\label{eq:R_SA}
\end{eqnarray}

\begin{figure}
\includegraphics[width=165mm]{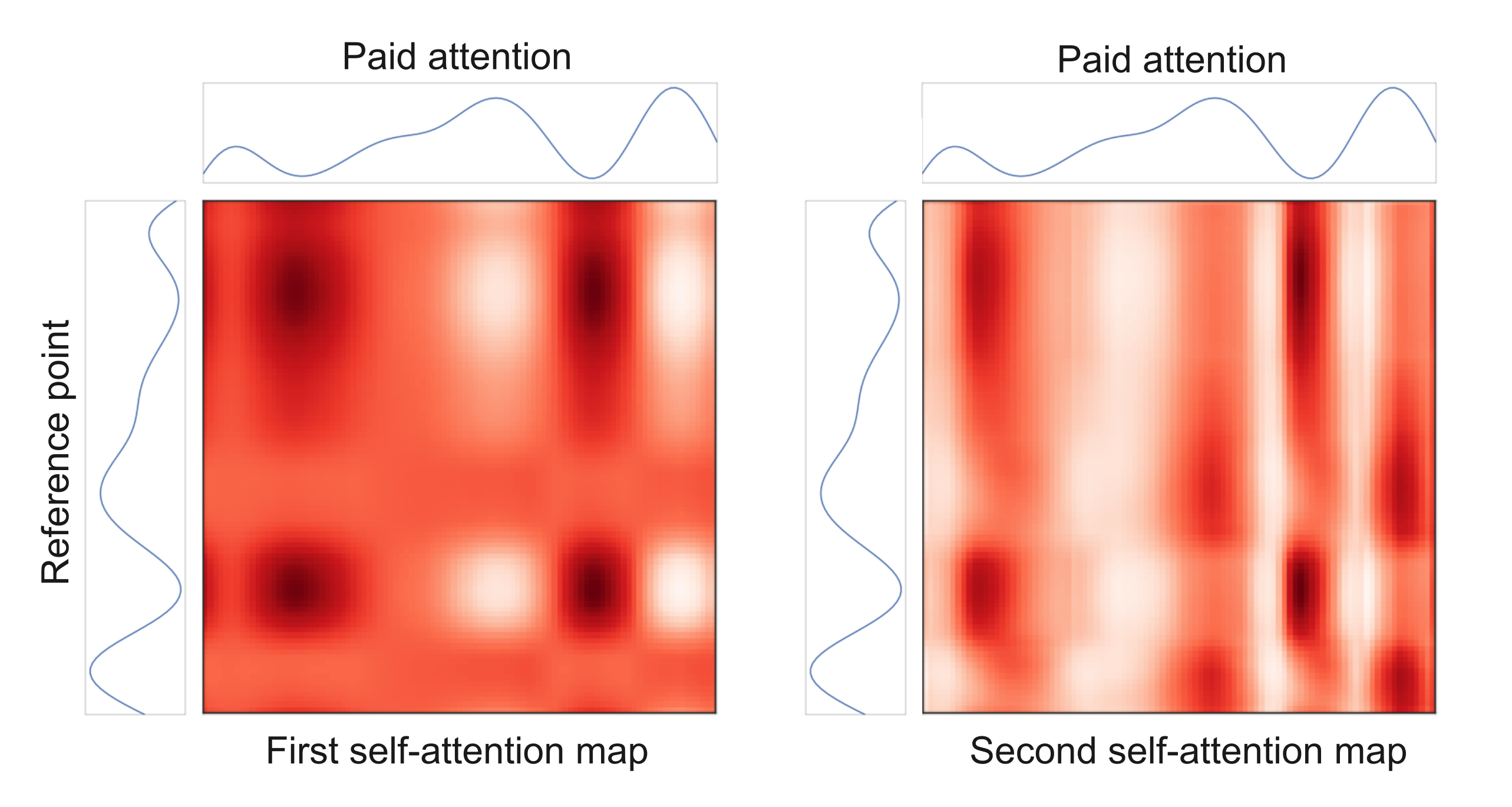}
\caption{\label{fig:attention_map}Attention maps for the potential in Fig.\ref{fig:learning_result}(c). The figure on the left shows the attention map of the first self-attention layer in the NN. The top and left sides of the figure show the corresponding potential shape. We can see which potential points the NN considers relevant for each reference point on the left side. The NN is intensely interested in the relationship between the minima of the potentials. The figure on the right shows the attention map of the NN's second self-attention layer. In this attention map, we can see that the NN is interested in the relationship between minima and maxima.}
\end{figure}

The quantity, $R_{SA}$, represents how much the loss is reduced by the self-attention mechanism. As shown in the right panel of Fig.\ref{fig:resnet_sa_effect}, the self-attention mechanism reduces losses for all training and validation dataset combinations. It is especially effective when smooth potentials are used for validation. We examine the attention maps to determine what the self-attention mechanism is focusing on.   Figure \ref{fig:attention_map} depicts the first and second self-attention maps for the potential depicted in Fig.\ref{fig:generalization}(c).  We used the NN trained with N5 dataset generated from the exponential kernel. To make the figure easier to understand, we aligned the potential shapes on the left and top sides of the self-attention maps. This self-attention map represents the importance of the relationship between two points on the potential learned by the NN. The figure on the left is the self-attention map placed in the layer closest to the input data and shows that the NN perceived the relationship between two minima of the potential as essential. The figure on the right is a self-attention map placed in the second self-attention layer. The NN perceived the relationship of maxima locations as meaningful in addition to the relationship of minima locations. Thus, attention to the relationship between distant feature points seems to reduce losses.  This observation also shows that the NN learns the necessary features by itself through learning.

The findings in this subsection show that NNs can help physics experiments produce more results with fewer practical resources.  For example, even when only materials with rough surfaces are available, it would be possible to train an NN with the results of experiments using those materials and then use the trained NN to predict the results of experiments using materials with smooth surfaces.  In addition, for a physical system with wide parameter space, we can experiment with a part of it and use the observed results to train the NN.  Then, we may raise candidates for parameter regions where interesting physical phenomena would be observed by using the trained NN to search for unknown parameter regions.  A schematic diagram is shown in Fig.\ref{fig:application}.

\subsection{\label{subsec:potential}Elicit information that is not being fed to the neural network}
When we trained the NN, we only gave it the potential and the corresponding energy eigenvalues. However, if the NN understands the physics described by Eq.(\ref{sch-eq}) in this process, we may extract a physical quantity other than the energy eigenvalues. This subsection attempts to find the probability of the particle’s existence, which is not used in training.

The function $f_0(v_m;\{w_{\alpha}\})$  of the learned NN predicts the energy eigenvalues of the ground state using values at each point of the potential. If $f_0(v_m;\{w_{\alpha}\})$ sensitively changes when the value of the potential at a point, $v_m$, is changed, there is a high probability that there is a particle in the vicinity. Therefore, the first-order derivative of $f_0(v_m;\{w_{\alpha}\})$ to the potential value is considered the existence probability of the particle for the ground state. The same can be said for excited states.

This can also be explained, as follows: $f_i(v_m;\{w_{\alpha}\})$ should have been learned to approximate $E_i$'s representation. From the discretized quantum mechanics, Eq.(\ref{eq:equation_of_motion})-(\ref{eq:wave-nomalization}), the first-order derivative of the energy eigenvalue to the potential value is the square of the eigenvector, as shown below,
\begin{eqnarray}
\frac{d E_i}{d v_m}
=
\frac{d}{d v_m}\left(\Psi^{T}_i H\Psi_i\right)
=
\Psi^{T}_i\left(\frac{d}{d v_m} H\right)\Psi_i+E_i\frac{d}{d v_m}\left(\Psi^{T}_i\Psi_i\right)
=
\psi^2_i(x_m)\label{eq:dedv}.
\end{eqnarray}
This should result in the following relationship if the learning is sufficient:
\begin{eqnarray}
\left.\frac{\partial f_i(v_m;\{w_{\alpha}\})}{\partial v_m}\right|_{\{w_{\alpha}\}={\rm learned\ values}}
\simeq
\psi^2_i(x_m)\label{eq:dfdv}.
\end{eqnarray}

As a property of NNs, the first-order derivative of the argument can be easily obtained using back-propagation. When training an NN, only the derivative to the parameters $\{w_{\alpha}\}$ is used, not the derivative to the input data $v_m$, but, here, we reverse the position and fix the parameters $\{w_{\alpha}\}$ on the trained values and obtain the derivative to the input data $v_m$.

\begin{figure}
\includegraphics[width=165mm]{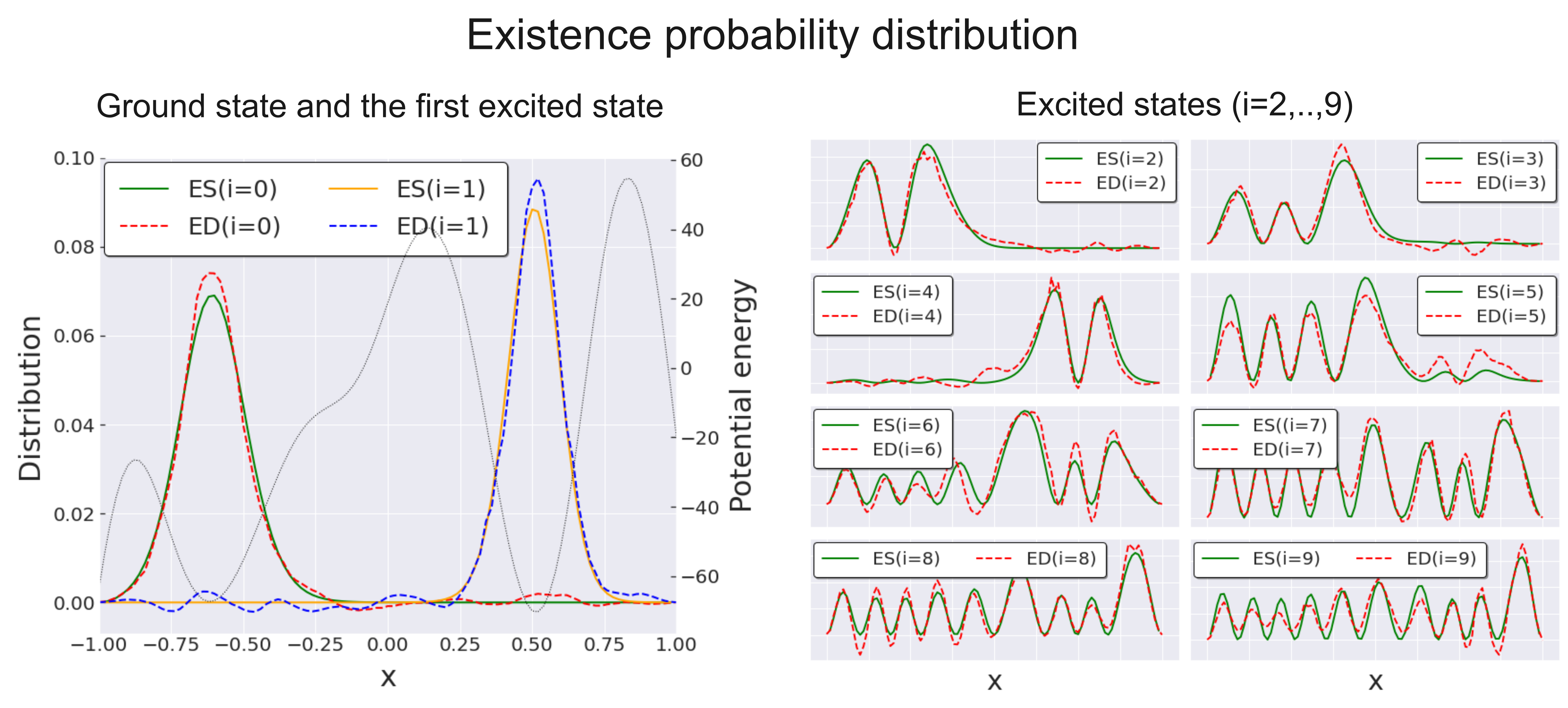}
\caption{\label{fig:distribution}
A neural network estimated the probability distribution of particle existence. For the ground state and the first excited state, both sides of Eq.(\ref{eq:dfdv}), the eigenvector squared (ES) and the predicted energy derivative (ED), are shown on the left. The same quantities are displayed in the figure on the right for an even higher excited state.
}
\end{figure}

The left and right sides of Eq.(\ref{eq:dfdv}) are displayed in Fig.\ref{fig:distribution}. The NN was trained with N5 dataset generated from the exponential kernel. The derivative values were not normalized again, and the values were displayed as they were. Although there are some unnatural parts, such as negative values, the general agreement regarding the location of the peaks and the number of nodes is excellent. This result of obtaining the probability distribution of the particle’s existence is surprising, considering that the used NN is only given the jagged potential such as Fig.\ref{fig:potential}(d) and the corresponding energy eigenvalues.

This subsection shows that the trained NN provides observable quantity other than the output data used during training. As a technical aspect, we also found that differentiation on input data using back-propagation may be helpful in physics. The same technic called saliency map has been used in object detection\cite{simonyan2013deep}. For example, by differentiating the dog probability in the outputs concerning the input image, one can get an idea of where a dog is in an image. It is very similar to the calculation in this subsection in that it focuses on the response to the input perturbation.

\subsection{\label{subsec:crossl}Predicting Physical Phenomena with Neural Networks}
We reproduce a well-known physical phenomenon here, but the same procedure would be able to predict unknown phenomena. 1D quantum mechanics has no degeneracy of energy eigenvalues, as can be seen from the fact that even the symmetric double-well potential has no degeneracy of energy eigenvalues. Two energy eigenvalues do not intersect, and their eigenstates cross over when the potential changes slowly and continuously. This phenomenon plays an essential role in the Mikheyev-Smirnov-Wolfenstein effect of the solar neutrino problem\cite{wolfensteinNeutrino1978,mikheyev1986resonant,parkeNonadiabatic1986,bethePossible1986}.

\begin{figure}
\includegraphics[width=165mm]{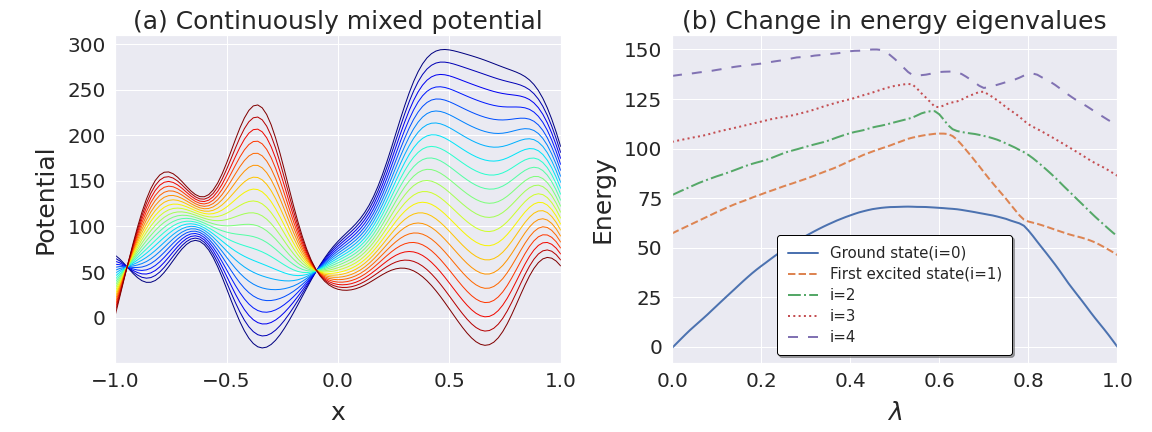}
\caption{\label{fig:level_crossing}Continuously varying potential and corresponding energy eigenvalue changes. (a) Two Gaussian kernel-generated potentials adjusted to zero ground state energy were linearly combined with a mixing constant $\lambda$. (b) An NN trained with the dataset generated from the exponential kernel was used to predict the evolution of the energy eigenvalues of the potentials generated in (a). The energy eigenvalues are changing without degenerating.}
\end{figure}

The following setup is used to reproduce this phenomenon. First, we generate two potentials, $v^{(1)}(x)$ and $v^{(2)}(x)$, using the Gaussian kernel, and subtract the energy eigenvalues of the respective ground states to create potentials $\tilde{v}^{(1)}(x)$ and $\tilde{v}^{(2)}(x)$ adjusted so that the energy eigenvalues of the ground states become zero. We then introduce a real mixing parameter $\lambda\in [0,1]$ and create a potential $v(x)$ that varies continuously from the linear combination, as follows:
\begin{equation}
    v(x)=(1-\lambda)\tilde{v}^{(1)}(x)+\lambda\tilde{v}^{(2)}(x).\label{eq:continuous_potential}
\end{equation}
This continuously varying potential is shown in Fig.\ref{fig:level_crossing}(a).

\begin{figure}
\includegraphics[width=165mm]{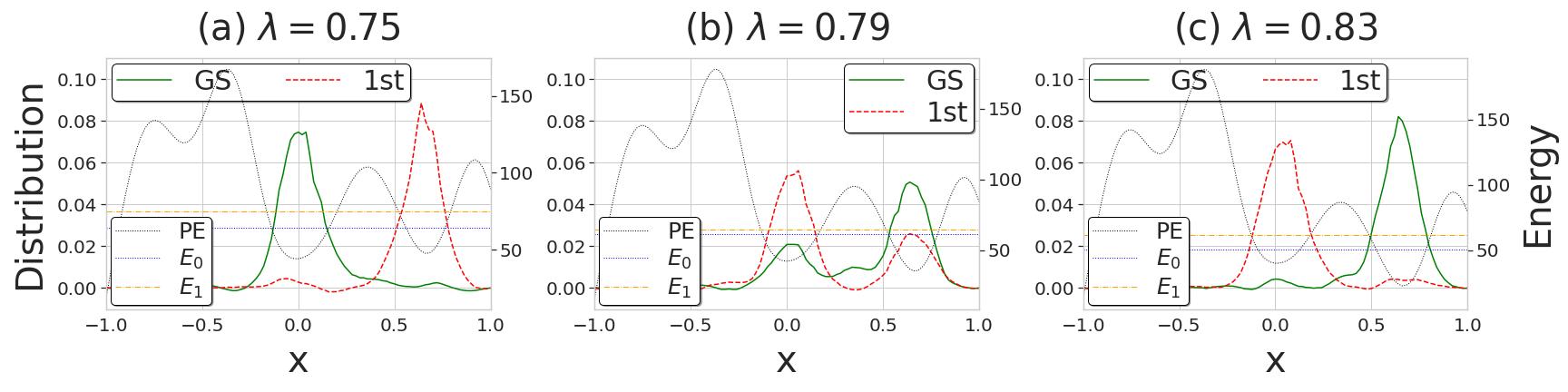}
\caption{\label{fig:cross_over}Crossing over between the ground state(GS) and the first excited state(1st) for the changing potential energy(PE). (a) The probability of the existence of a particle at $\lambda = 0.75$. (b) Existence probability of the particle at $\lambda = 0.79$.(c) Existence probability of the particle at $\lambda = 0.83$. We used the left side of Eq.(\ref{eq:dfdv}) for this estimation.}
\end{figure}

Figure \ref{fig:level_crossing}(b) shows the predicted energy eigenvalues from the NN as a function of $\lambda$. The energy eigenvalues are changing without crossing. The NN trained with the N5 dataset generated from the exponential kernel was used. The energy eigenvalues of the ground state and the energy eigenvalue of the first excited state are comparable around $\lambda = 0.8$. We investigate what is happening in the vicinity of this value using the method of Sec.\ref{subsec:potential}. Figure \ref{fig:cross_over}(a) shows the existence probability distribution of the particle at $\lambda = 0.75$. The ground state is on the left and the first excited state is on the right. Figure \ref{fig:cross_over}(b) shows the existence probability distribution of the particle at $\lambda = 0.79$. The existence probability distributions of the ground state and the first excited state overlap. Figure \ref{fig:cross_over}(c) shows the existence probability distribution of the particle at $\lambda = 0.83$. The ground state is on the right, and the first excited state is on the left. In this way, we can see how the eigenstates continuously cross over.

This subsection showed that the trained NN could reproduce the well-known physical phenomenon. Therefore, it is also expected to help humans to find new physical phenomena.

\subsection{\label{subsec:adaptability}Wide range of applications and experiments necessary for learning}
So far, we've looked at the case where the experimenter is fully aware of the potential. We also know that the physical system was governed by the Schrödinger equation, so we can solve it numerically and obtain the energy eigenvalues in this case. In this subsection, we consider simple toy models to demonstrate that the NN method has a broader applicability range.

Consider the case of an experiment with a particle in the matter. The experimenters can make tunable potentials, $V_e(x)$, to the particle by employing an electric field. However, the potentials are expected to be $V_{total}(x)$ in matter due to unknown matter effects to the experimenters. Because the experimenters do not know how matter affects the potential in this case, the Schrödinger equation cannot be used to determine the energy eigenvalues. Even in this case, if they can observe the energy eigenvalues for various $V_e(x)$ in the experiment, they can train NN with the results and predict the energy eigenvalues using the trained NN without knowing the laws governing the system.

The matter effects used in Sec.\ref{sec:case1} and Sec.\ref{sec:case2} are simple examples and do not model specific physical systems. We employ simple matter effects of comparable magnitude to $V_e(x)$. However, the method used here is sufficiently general that it works independently of the choice of the matter effect function.

\subsubsection{\label{sec:case1}Case 1}
First, as a simple case, consider that a fixed potential is added as follows($\sigma = 100$),
\begin{equation}
    V_{total}(x)=V_e(x)+\sigma \sin{(\pi x)}.\label{eq:case1eq}
\end{equation}
We assume the particle moves in $-1<x<1$, and the experimenters can vary $V_e(x)$ in a variety of ways. The experimenters do not know the matter effect, $\sigma \sin{(\pi x)}$, and then $V_{total}(x)$, but the energy eigenvalues are assumed to be observable in the experiment. The experimenters can train NN with their generated $V_e(x)$ as inputs and the observed energy eigenvalues as outputs.

We here use the N5 dataset as the input potentials, $V_{e}(x)$, for the numerical experiment. The energy eigenvalues used for the ground truth data are calculated from $V_{total}(x)$ using Schrödinger equation to make artificial experimental results datasets.  These values correspond to the observed energy eigenvalues without the experimental error.

\begin{figure}
\includegraphics[width=165mm]{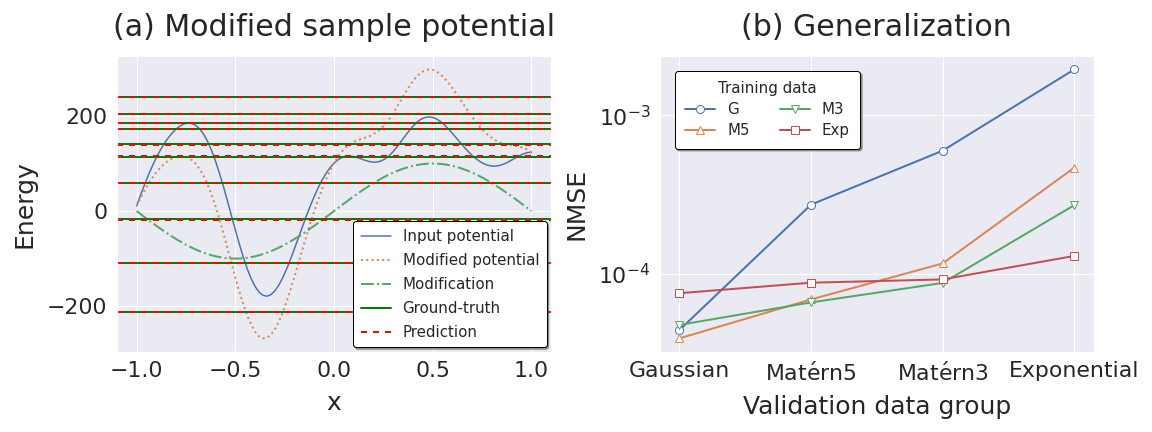}
\caption{\label{fig:sin}
(a) The energy eigenvalues predicted by NN for a modified potential by the law, Eq.(\ref{eq:case1eq}).  The N5 exponential dataset is used for training. 
(b) NMSE for the combinations of the training and validation datasets in case 1.
}
\end{figure}

Figure \ref{fig:sin} shows the results of numerical experiments in this case. The left figure shows the input potentials $V_e(x)$, $\sigma \sin{(\pi x)}$, $V_{total}(x)$, the ground truth energy eigenvalues, and their predictions. This figure shows that the energy eigenvalues are correctly predicted. The NMSE is depicted in the figure on the right. The losses are comparable to those shown in Fig.\ref{fig:generalization}(b), indicating that the learning has progressed sufficiently. These figures demonstrate that NN comprehends the law composed of Eqs. (\ref{sch-eq}) and (\ref{eq:case1eq}).

\subsubsection{\label{sec:case2}Case 2}
Next, consider more complex case where the matter effect depends on $V_e(x)$ as follows,
\begin{equation}
    V_{total}(x)=V_e(x)+\frac{1}{2\sigma}V_e(x)^2-\sigma.\label{eq:case2eq}
\end{equation}
We also use the N5 dataset as the input potential, $V_{e}(x)$, and $\sigma = 100$.

\begin{figure}
\includegraphics[width=165mm]{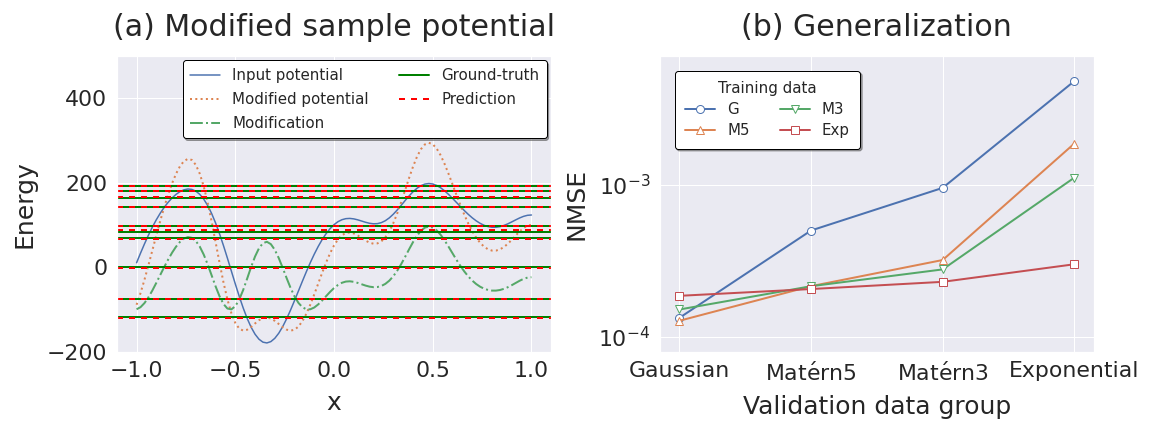}
\caption{\label{fig:quad}
(a) The energy eigenvalues predicted by NN for a modified potential by the law, Eq.(\ref{eq:case2eq}).  The N5 exponential dataset is used for training. 
(b) NMSE for the combinations of the training and validation dataset in case 2.
}
\end{figure}

Figure \ref{fig:quad} shows the results of numerical experiments in this case. The left figure shows the input potentials $V_e(x)$, $\frac{1}{2\sigma}V_e(x)^2-\sigma$, $V_{total}(x)$, the ground truth energy eigenvalues, and their predictions. This figure shows that the energy eigenvalues are also correctly predicted in this case. This figure shows that the energy eigenvalues are also correctly predicted in this case. The NMSE is depicted in the figure on the right. In comparison to Fig.\ref{fig:sin}(b) in Case 1, the error is slightly larger, but learning has progressed sufficiently in this case as well. This diagram shows that NN comprehends the law composed of Eqs.(\ref{sch-eq}) and (\ref{eq:case2eq}).

Even though $V_{total}$ is not given, the NN can predict the ten energy eigenvalues for given $V_e$ in both cases, 1 and 2. These numerical experiments demonstrate that the NN method can be used even when the laws governing the physical system are unknown, as long as sufficient experimental results are available.

\subsubsection{\label{sec:experiment}Necessary experiments and NN's versatility}
As we have seen from the numerical experiments in cases 1 and 2, the NN can be trained to make predictions even if the experimenters do not know the laws governing the physical system.  However, it is necessary to provide a variety of inputs and obtain the corresponding experimental results as outputs. As a result, an experimental setup should be prepared to automatically generate various inputs while monitoring the results.

As shown in Fig.\ref{fig:sin}(b) and Fig.\ref{fig:quad}(b), the NN trained on the potentials generated by the exponential kernel predicts well the potentials generated by the other kernels. This finding suggests that if NNs are trained on a variety of easily experimented-with inputs, they can predict the outcomes of experiments even in difficult-to-experiment-with input domains. This observation demonstrates that sufficiently versatile NNs can be created even from experiments in a limited environment.

This NN-based method would also be appropriate when the laws governing a physical system are known but difficult to solve numerically, due to high dimension or multi-particle interactions, for example. Even in such cases, NN can be trained using a variety of experimental results. Once the trained NN is obtained, the results of experiments can be predicted.

\section{\label{sec:conclusion}Conclusion}
This paper investigated how NNs understand physics using 1D quantum mechanics as a subject. In the preparation stage in Sec.\ref{sec:method}, it was found that a DNN can find energy eigenvalues only from the potentials with errors of approximately 1$\%$.

We showed from four aspects that NNs could understand physics and are helpful for physics research in Sec.\ref{sec:understanding}.
\begin{itemize}
    \item 
    The NN could predict energy eigenvalues, even for different kinds of potentials from those used for training. For example, a NN trained on a dataset created from jagged potentials could successfully predict the energy eigenvalues of a smooth potential. We also found that capturing the overall potential shape using the self-attention mechanism and deepening the NN through residual-connection improved prediction accuracy.
    \item
    The NN could predict the existence probability distribution of particles not used during training. The results show that extracting physical quantities not given during training from the NN is possible.
    \item
    The NN predicted the crossover of states when the potential changes slowly and continuously. The results show that NN can reproduce or predict physical phenomena.
    \item
    NNs are found to be trained with diverse inputs and the corresponding experimental results, even when the laws governing the physical system are unknown to the experimenters. Once the NN has been trained, it can be used to predict the experimental results of the systems.
\end{itemize}

\begin{figure}
\includegraphics[width=165mm]{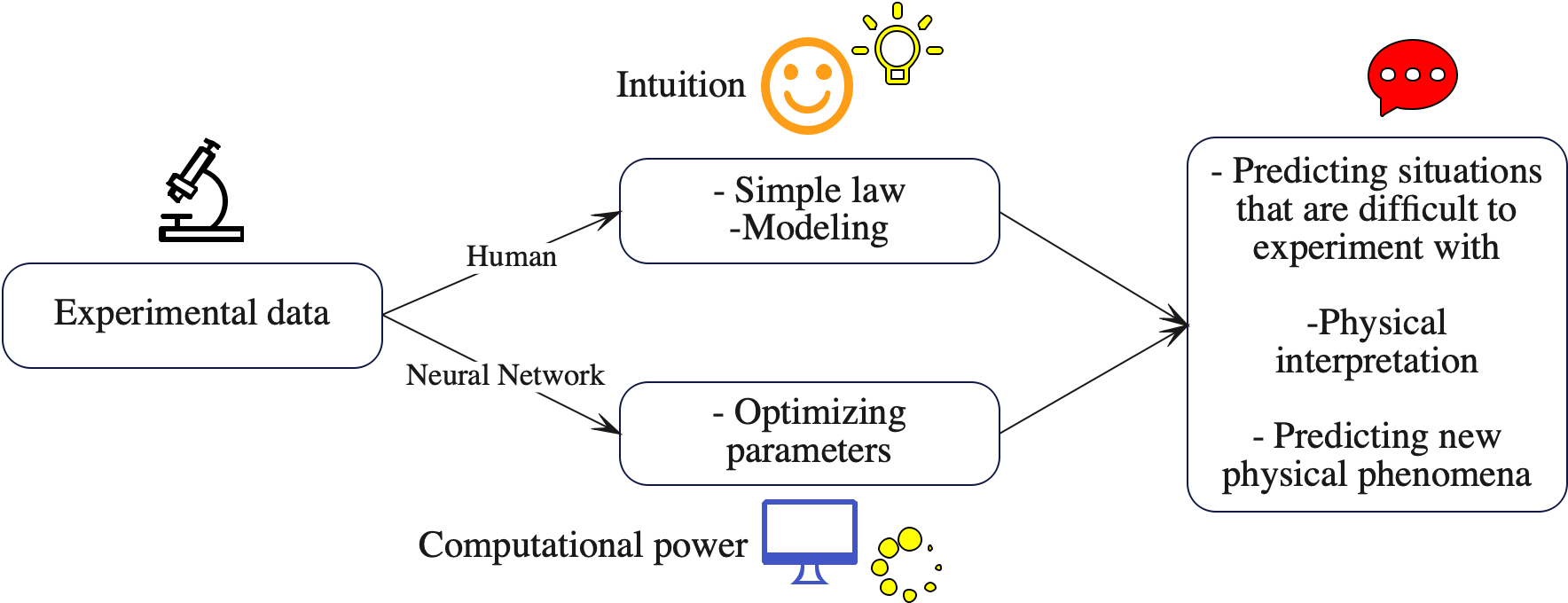}
\caption{\label{fig:interpretation}Differences in the understanding of physics by humans and neural networks. Humans look at the results of physical experiments and use their intuition to derive concise physical laws and create models that describe physical systems well. On the other hand, neural networks use significant computational power to optimize numerous parameters from the results of physical experiments. Both can make predictions about physics, but the process is entirely different.}
\end{figure}

As shown in Fig.\ref{fig:interpretation}, how NNs and humans understand physics seems different. Humans understand physics using their intuition from experimental data, discover unifying laws such as the Schr\"{o}dinger equation that governs physical systems, or build phenomenological approximation models. On the other hand, NNs attempt to understand physical systems by adjusting numerous parameters, taking advantage of their high computational power to explain a large amount of experimental data. Although humans and NNs follow different paths to understand physics, the underlying physics is the same, and, as a result, they can make the same predictions.

\begin{figure}
\begin{center}
\includegraphics[width=120mm]{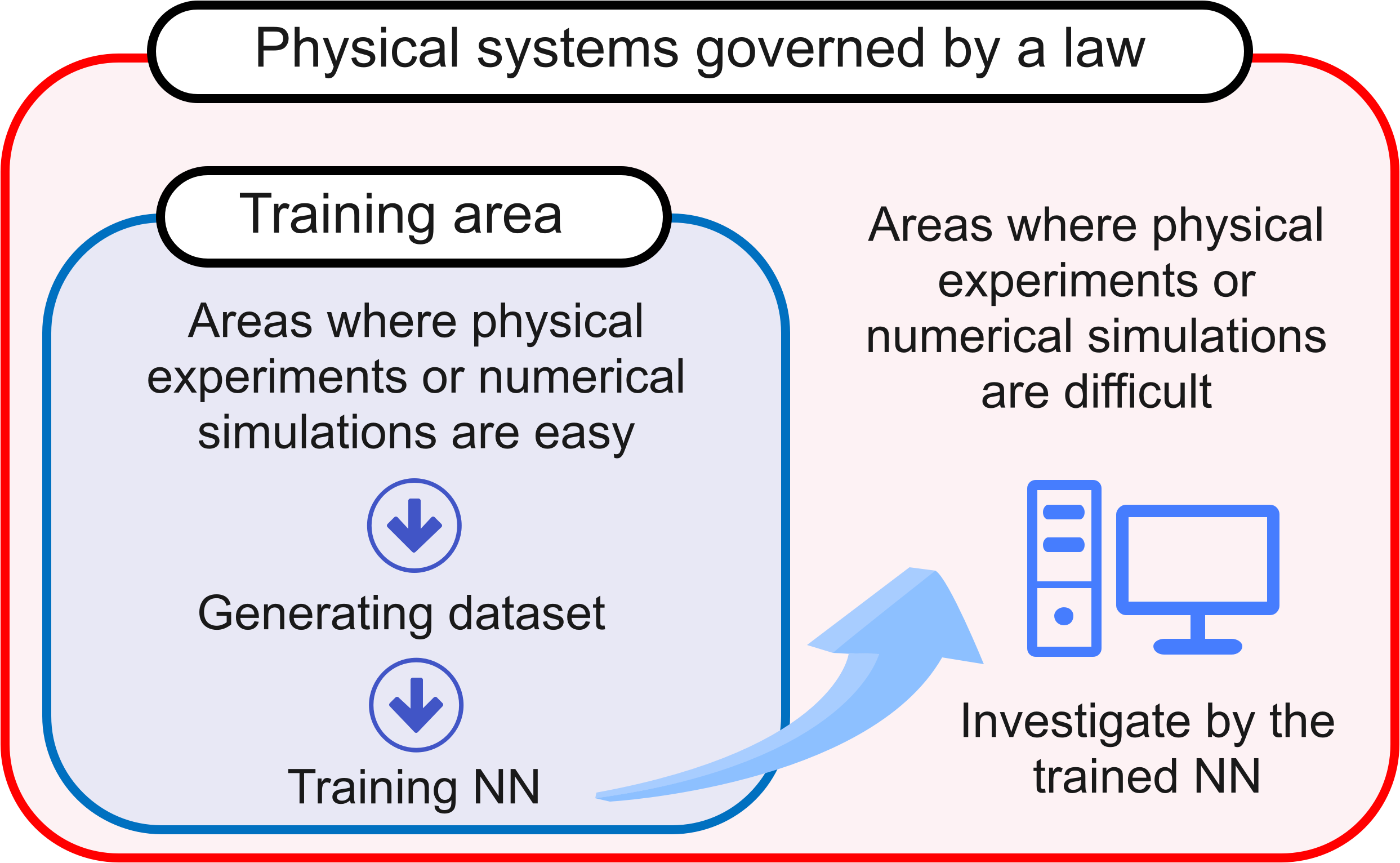}
\caption{\label{fig:application}
Using NN for physics research. Initially, a dataset is prepared in a training area within a physical system that adheres to the same physical laws. If the dataset is created through experimentation, this training area will meet conditions such as temperature and material, making it easy to conduct experiments. If the dataset was generated using a numerical simulation, such as a Monte Carlo method, the training area corresponds to a parameter region with fast convergence. NNs can be trained using data generated by experiments or numerical simulations. As demonstrated in this paper, NNs are versatile and can be used to predict experimental results under difficult conditions to perform experiments or numerical simulations.
}
\end{center}
\end{figure}

The understanding of physical systems by NNs has the following advantages.
\begin{itemize}
    \item  
    It allows us to tackle problems without prior intuition. The NN used in this paper is a combination of layers used in natural language processing and image recognition, and no part of it has been specially devised for physics.
    \item
    We can create a dataset in a region where experiments or numerical simulations are simple to perform and use it to train NNs. As demonstrated in this paper, this trained NN is versatile and can be used to make predictions in conditions that are difficult to experiment with or numerically simulate. Figure \ref{fig:application} depicts a schematic diagram of the procedure. 
\end{itemize}

On the other hand, it has the following drawbacks.
\begin{itemize}
    \item  
    Massive amounts of experimental data or highly diverse numerical simulation results are required to train a NN with high reproducibility while avoiding overfitting. Experiments for these preparations must be automated. High computer performance and clever algorithms are required when numerical simulations, such as Monte Carlo methods, generate the training dataset.
    \item 
   It is possible to improve the approximation's accuracy, but it does not appear to be as simple to obtain the exact solution as the Schrödinger equation so far. Humans still appear to be in charge of finding exact solutions or approximating physical systems with a model with few parameters. However, the ability of NNs to predict the outcomes of experiments under a variety of conditions will aid humans in locating them. 
\end{itemize}

These findings demonstrate that NNs can learn physical laws from experimental data, predict experimental results under conditions different from those used for training, and predict physical quantities of types not provided during training. Because NNs understand physics in a different way than humans, they will be a powerful tool for advancing physics by complementing the human way of understanding. 

\appendix
\section{\label{sec:architecture}Architecture of Neural Networks}
This appendix describes the architecture of the NN we used in this paper. Our NN consists of a combination of the Resnet and SAGAN structures, which have recently been used with success in natural language processing and image generation\cite{he2016deep,zhang2019self}.
 The residual connection used in Resnet mitigates gradient vanishing and increases the depth of the NN. On the other hand, the self-attention mechanism used in SAGAN mitigates the shortcoming that convolutional layers cannot relate to distant neurons and is very powerful in natural language processing and image generation.

\begin{figure}
\includegraphics[width=165mm]{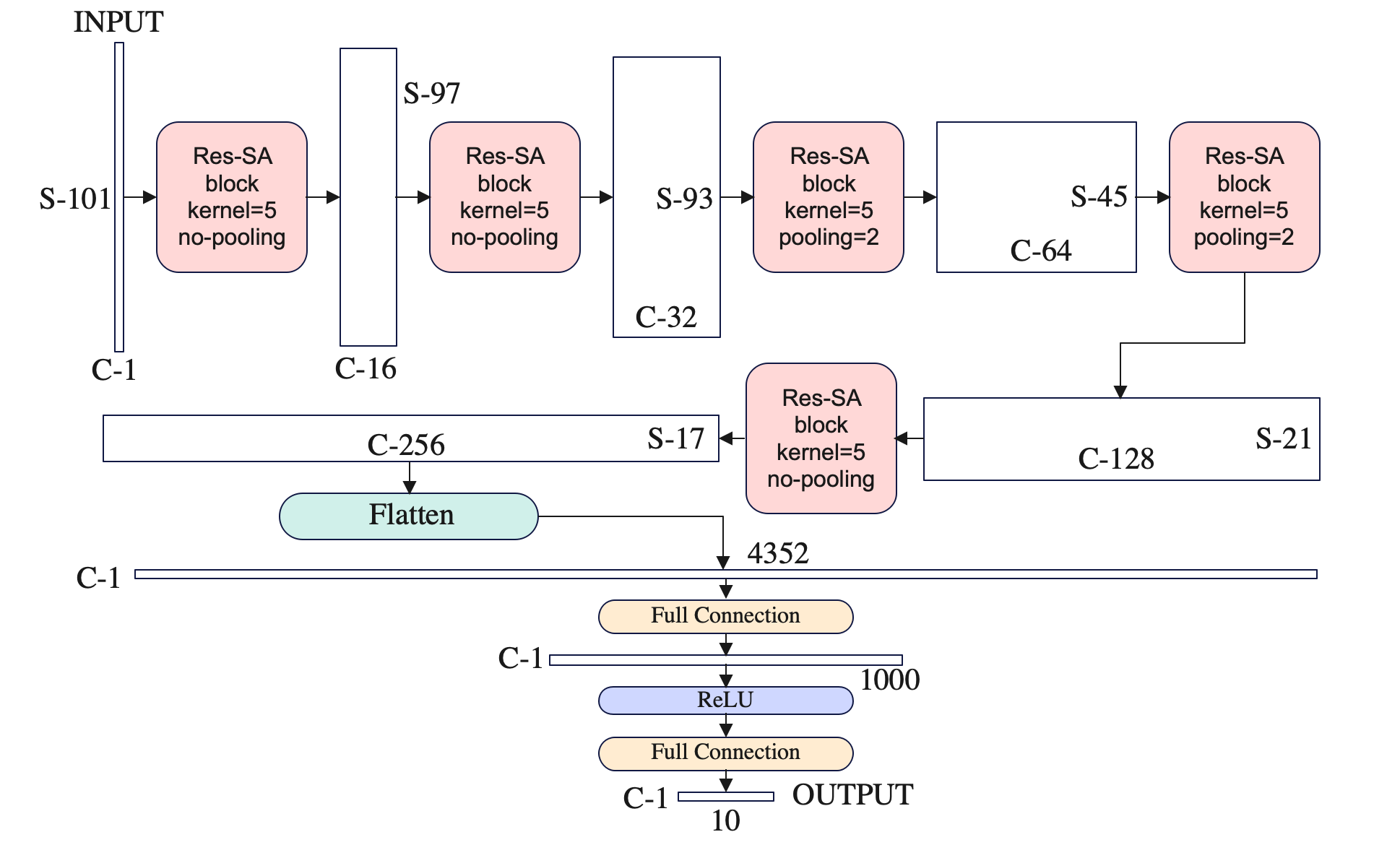}
\caption{\label{fig:whole}Design of the entire neural network. The rectangles show the shapes of the tensor at that position. First, the input is the potential value at 101 points in a one-dimensional space. Then, passing through the Res-SA block five times, the tensor becomes longer in the channel direction and shorter in the spatial direction. Finally, it is transformed into a flattened 1-dimensional tensor, which goes through full connection layers and becomes a 10-dimensional vector. This 10-dimensional vector corresponds to the values of the ten lowest energy eigenvalues. The Res-SA block in this neural network is depicted in detail in Fig.\ref{fig:res-sa}.}
\end{figure}

We began with the overall structure, as shown in Fig.\ref{fig:whole}. First, the input is a 101-dimensional vector corresponding to the potential value at each point. Next, this input is stretched in the channel direction by the blocks containing the convolutional layer, reducing it to 17 dimensions in the spatial direction, whereas it becomes 256 dimensions in the channel direction. Finally, it is flattened to a 1D vector and passed through full connection layers, outputting a 10-dimensional vector. This 10-dimensional vector corresponds to ten small energy eigenvalues.

\begin{figure}
\includegraphics[width=165mm]{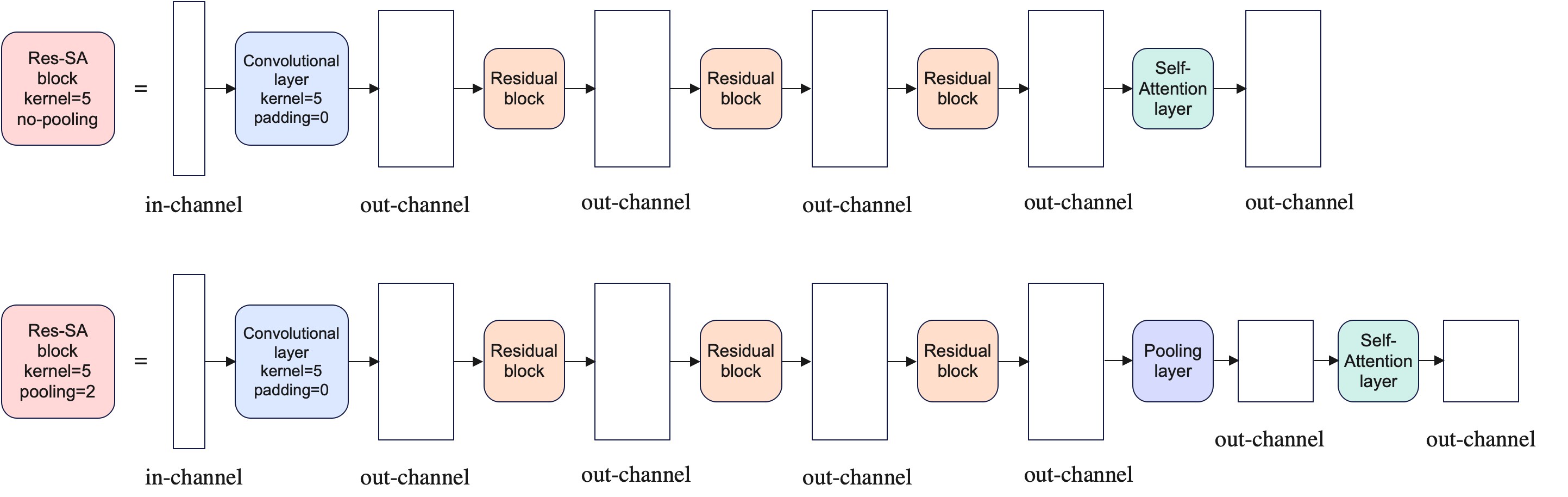}
\caption{\label{fig:res-sa}Res-SA block design. The input tensor can change the number of channels in the initial convolution layer. At that time, the kernel size is 5, and the padding size is 0, so the tensor size in the spatial direction is reduced by 4. If there is a pooling layer, the tensor size in the spatial direction is halved there. For the other layers, the size of the tensor does not change. The deep structure increases the flexibility of the function, and the self-attention layer helps relate the information of points far apart in position. For the self-attention layer, we used the same structure as in the SAGAN paper\cite{zhang2019self}. The details of the residual block are depicted in Fig.\ref{fig:residual}.}
\end{figure}

Figure \ref{fig:res-sa} shows the details of the Res-SA block used in Fig.\ref{fig:whole}. The input first passes through the convolutional layer and then passes through the residual block three times. Finally, a self-attention layer is used to relate the information about distant locations. We embedded the self-attention layer in the same way as in the SAGAN study\cite{zhang2019self}. The only difference is that SAGAN needed to flatten a 2D image to 1D to create the attention map, but our 1D potential can skip this process.

\begin{figure}
\includegraphics[width=165mm]{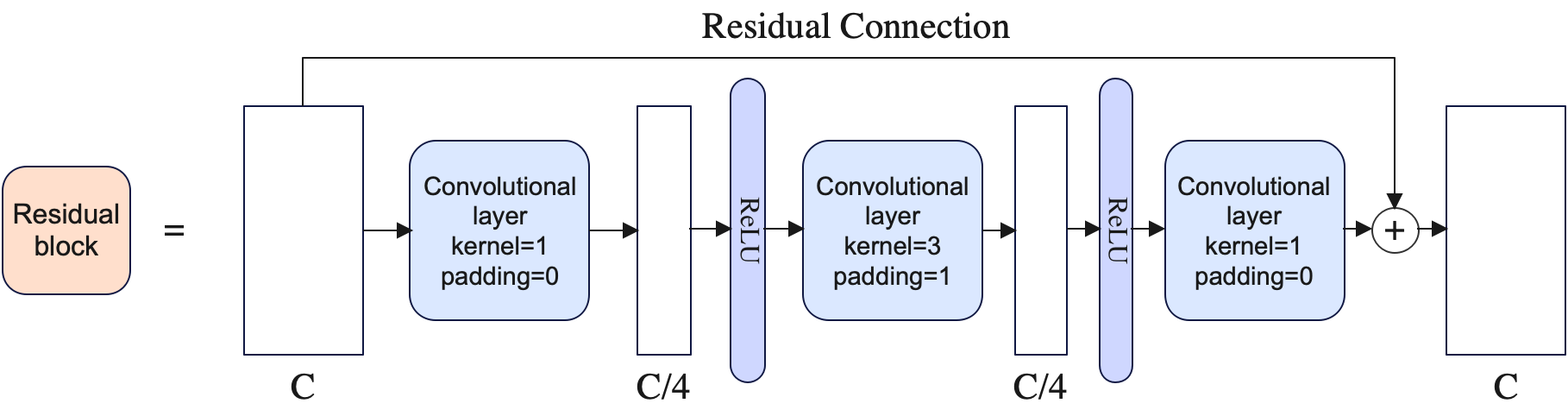}
\caption{\label{fig:residual} Design of the residual block. The block consists of two pointwise convolutional layers corresponding to full connection layers in the channel direction and a convolutional layer with a kernel size of 3. The padding is adjusted so that the input and output tensors have the same shape. This structure is the same as the one used in Resnet. The presence of residual coupling allows learning to proceed without gradient loss, even as the depth of the neural network increases.}
\end{figure}

In Fig.\ref{fig:residual}, we show the structure of the residual block. This block consists of three convolutional layers and one residual connection and has the feature that the input and output have the same shape. This structure is the same as the one used in Resnet.

We have employed the ReLU function for all activation functions, totaling 31 times. The total number of parameters used in this network is approximately $4.9\times 10^6$.

Next, we explain the specific method of learning. We employed Adam as the optimization method and varied the learning rate from $1\times 10^{-3}$ (initial value) to $1\times 10^{-5}$ (value at the last epoch) in a geometrical progression. In addition, since learning is more efficient when the input and output values are of order O(1), we multiplied the input and output values by 0.01 and adjusted them to be of that magnitude. The batch size was fixed to 100. Pytorch was the NN framework used, and the GPy library provided the kernel functions used. The Numpy library generated the random numbers from the covariance matrix. For diagonalization of the matrix, we used the
Scikit-learn library.

\bibliographystyle{ptephy}
\bibliography{quantum_mechanics,neural_network,book,physics}

\end{document}